\documentclass{elsart}
\usepackage{amssymb}
\usepackage{amsmath}
\usepackage[singlespacing]{setspace}

\input{tcilatex}
\begin{document}

\begin{frontmatter}

\title{Bifurcations in nonlinear models of fluid-conveying pipes supported at both ends}

\author{M. Nikoli\'c and M. Rajkovi\'c *}
\address{*milanr@vin.bg.ac.yu}
\address{Institute of Nuclear Sciences Vin\v ca,  
   P.O. Box 522, 11001 Belgrade, Serbia }

\begin{abstract}
Stationary bifurcations in several nonlinear models of fluid conveying pipes fixed at 
both ends are analyzed with the use of Lyapunov-Schmidt reduction and singularity
theory. Influence of gravitational force, curvature and vertical elastic support on 
various properties of bifurcating solutions are investigated. In particular the conditions 
for occurrence of supercritical and subcritical bifurcations are presented for the models of Holmes, 
Thurman and Mote, and  Paidoussis.   
\end{abstract}


\end{frontmatter}

\section*{Introduction}

The nonlinear dynamics behavior of pipes conveying fluid displays
interesting and paradigmatic behavior with important practical implications
and it offers a rich setting for development and testing of nonlinear
dynamics theory. Bifurcation theory represents one of the main subject areas
of nonlinear dynamics, analyzing behavior either in the vicinity of trivial
solutions (local bifurcations), or the existence in-the-large of connected
sets of non-trivial solutions (global bifurcations). In the study of local
bifurcations, particularly interesting is the influence of various
parameters figuring in the governing equations on the location and stability
of fixed points and on classification of bifurcations as supercritical or
subcritical. In a supercritical bifurcation there is no discontinuous change
in size and form of the attractor, and after the bifurcation the new
(enlarged) attractor contains within itself the old attractor. On the other
hand, in subcritical bifurcation the attractor disappears followed by a jump
of the system to a remote and completely new attractor, via a fast dynamic
transient. For the complete understanding of the dynamics, both stationary
and nonstationary (dynamic) aspects of the bifurcation theory are important.

In the case of fluid-conveying pipes, nonstationary analysis is related to
the study of supercritical and subcritical Hopf bifurcations, the
determination of the amplitude associated with flutter and the dependence of
oscillation frequency on the amplitude. Stationary analysis, on the other
hand, is concerned with stationary bifurcations i.e. changes in the
equilibrium point structure of the underlying equations, which due to
reflection symmetry are of pitchfork type. Furthermore interplay between
stationary and dynamic bifurcations may lead to more complicated dynamics
such as spatio-temporal intermittency and chaos (Argentina and Collet,
1998). In spite of many publications devoted to these subjects, a thorough
and systematic analysis of nonlinear models of fluid conveying pipes,
particularly of pipes supported at both ends, and related bifurcations is
still lacking. Following the pioneering works by Holmes (1977, 1978) further
work of interest appeared in Ch'ng (1977, 1979) and Lunn (1982). As a
paradigm of the studies performed so far, the work of Holmes based on the
center manifold analysis concentrated on one-equation model which does not
take into account any gravitational nor tensioning effects. The coefficients
figuring in the normal form of the obtained bifurcation equation were given
only as numerical values, disguising the influence of involved physical
quantities. Finally, the complete two-equation model of Paidoussis (2003)
has only recently been introduced in its correct form and it has never been
used. A wealth of up-to-date information related to the published works on
nonlinear aspects of fluid-conveying pipes with supported ends may be found
in Paidoussis (1998, 2003).

In order to improve upon the current status and in order to prepare the
ground for the study of spatio-temporal intermittency (and chaos), we study
several well-known and frequently used nonlinear models in which we focus on
nondegenerate local bifurcations from the trivial solution to another
stationary nontrivial solution. In particular, the object of our analysis
are pipes fixed at both ends shown in Fig. 1 and nonlinear models of Holmes
(1977), Thurman and Mote (1969) and the complete nonlinear model of
Paidoussis (2002), [the latter being based on the model of Semler at al.
(1994)]. The manner of presentation is such that each model is considered as
a special case of the complete nonlinear model, revealing the influence on
the complete dynamics of different linear and nonlinear terms and
quantities. Moreover the increasing complexity of each model leads to a
better understanding of the complete nonlinear model. Each model is
presented in a separate section of the paper, with subsections related to
certain important parameters influencing the dynamics, such as the
gravitational force or pipe curvature.

Since this exposition relies strongly on the Lyapunov-Schmidt reduction, a
brief description of the procedure is presented in Appendix A. The
interested reader may be find extensive treatment in books (Golubitsky et
al., 1985; Golubitsky and Schaeffer, 1988). An important characteristic of
the Lyapunov-Schmidt reduction is that the procedure for obtaining normal
form of the bifurcation equation is pursued by analytical means, without
using any numerical or truncating procedures. Hence the normal form is exact
and influence of each parameter on the type of bifurcation may be traced and
analyzed. An important insight gained by using this method is reflected in
the fact that we obtain exact analytical solutions in the vicinity of
bifurcation point for each model and derive exact conditions that classify
bifurcations as supercritical or subcritical.

\section{The Model of P. J. Holmes}

\subsection{Bifurcations in the stationary model without gravitational
effects}

Holmes considered pipes with supported, non-sliding ends and obtained a
nonlinear equation of motion by adding to a linear equation the nonlinear
term corresponding to the mean, deformation-induced tensioning. The complete
equation of motion in nondimensional form reads%
\begin{gather}
\alpha \overset{\cdot }{v}^{^{\prime \prime \prime \prime }}+v^{^{\prime
\prime \prime \prime }}-(\Gamma -\rho ^{2}-\frac{1}{2}\mathcal{A}\left\vert
v^{\prime }\right\vert ^{2}+\alpha \mathcal{A}\left\langle v^{\prime }\mid 
\overset{\cdot }{v}^{\prime }\right\rangle )v^{\prime \prime }  \label{1} \\
-\gamma ([1-\xi ]v^{\prime })^{\prime }+2\sqrt{\beta }\rho \overset{\cdot }{v%
}^{\prime }+\sigma \overset{\cdot }{v}+\text{ }\overset{\cdot \cdot }{v}%
\text{ }=0,  \notag
\end{gather}%
where $v=v(\xi ,t)$ denotes the lateral deflection normalized by the length
of the pipe. In addition, $\Gamma $ represents the tensile force on the
pipe, $\beta $ represents the mass ratio, $\rho $ the flow velocity, $%
\mathcal{A}$ the axial stiffness, $\alpha $ is related to the viscoelastic
structural damping, $\sigma $ represents fluid damping, and $\gamma $
denotes gravitational effects. Explicitly,%
\begin{equation}
\alpha =\left( \frac{EI}{M+m}\right) ^{1/2}\frac{a}{L^{2}},\text{ \ \ }\beta
=\frac{M}{M+m},\text{ \ \ }\gamma =\frac{M+m}{EI}L^{3}g,  \label{2}
\end{equation}%
where $a$ is the coefficient of Kelvin-Voigt damping in the pipe material, $%
M $ and $m$ are the mass of fluid and pipe per unit length, respectively,
and $L$ is the length of the pipe, while%
\begin{equation}
\Gamma =\frac{T_{0}L^{2}}{EI},\text{ \ }\mathcal{A}=\frac{EAL^{2}}{EI},\text{
\ \ }\sigma =\frac{cL^{2}}{[EI(M+m)]^{1/2}}.  \label{3}
\end{equation}%
In the above equations $T_{0},$ $EI,$ $L,$ $A,$ $I,$ $E$ and $c$ are the
longitudinal externally applied tension, flexural rigidity of the pipe,
length of the pipe, cross-sectional area, area-moment of inertia, modulus of
elasticity and constant of damping (due to friction), respectively. Moreover 
$v^{\prime }=dv/d\xi ,$ $\overset{\cdot }{v}=dv/dt.$ With $\xi =s/L,$ $%
\left\vert v\right\vert $ denotes the $L^{2}$ norm 
\begin{equation}
\left\vert v^{\prime }\right\vert =(\int\nolimits_{0}^{1}v^{\prime 2}(\xi
)d\xi )^{1/2},  \label{4}
\end{equation}%
while the $\left\langle v^{\prime }\mid \overset{\cdot }{v}^{\prime
}\right\rangle $ denotes the $L^{2}$ inner product 
\begin{equation}
\left\langle v^{\prime }\mid \overset{\cdot }{v}^{\prime }\right\rangle
=\int\nolimits_{0}^{1}v^{\prime }(\xi )\overset{\cdot }{v}^{\prime }(\xi
)d\xi .  \label{5}
\end{equation}%
The boundary conditions are $v=v^{\prime \prime }=0$ at $\xi =0,1$. The
corresponding stationary equation, neglecting gravitational effects, applied
tension and nonlinear dissipative term following a Lyapunov-Schmidt
reduction, locally may be put into one-to-one correspondence with solutions
of a single algebraic equation%
\begin{equation}
g(\mu ,\rho )=\left( \frac{3}{4}(n\pi )^{4}\mathcal{A}\right) \text{ }\mu
^{3}-(n\pi )^{3}\text{ }\mu \rho +\vartheta (3)=0,  \label{6}
\end{equation}%
indicating that the corresponding bifurcation is always a supercritical
pitchfork one, since%
\begin{equation*}
\frac{3}{4}(n\pi )^{4}\mathcal{A}>0.
\end{equation*}%
\qquad \qquad

Using a two-mode Galerkin discretization applied to the complete equation (%
\ref{1}) and assuming $\sigma $ $=g=\Gamma =0,$ Holmes (1977) obtained
numerical values for terms multiplying $\mathcal{A}\mu ^{3}$ and $\mu \rho $
respectively in the normal form equation (\ref{6}). We have shown here that
the sign of these terms is unalterable and we have presented their
analytical form. Applying the Lyapunov-Schmidt reduction to the same
stationary equation and taking applied tension into account, the following
normal form of the bifurcation equation is obtained:%
\begin{equation}
g(\mu ,\rho )=\frac{3}{4}(n\pi +\frac{\Gamma }{2n\pi })^{4}\mathcal{A}\left[
1-\frac{\sin ^{2}\left( \frac{\Gamma }{n\pi }\right) }{4(n\pi +\frac{\Gamma 
}{2n\pi })^{2}}\right] \text{ }\mu ^{3}-(n\pi )^{3}\text{ }\mu \rho
+\vartheta (3)=0.  \label{7}
\end{equation}%
In this case too, the type of pitchfork bifurcation is supercritical and may
not be altered, irrespective of the value of tension $\Gamma $ or axial
flexibility\ $\mathcal{A}$.\ 

Since the influence of gravitational effects was not considered in the
original treatments (Holmes, 1977, 1978), nor in subsequent studies, we
analyze this effect along with the effect of tensioning in the next section,
again restricting analysis to the stationary version of equation (\ref{1}).

\subsection{The role of gravitational effects}

The stationary equation representing nonlinear dynamic behavior of a pipe
conveying fluid including gravitational and tensioning effects may be
written in the form%
\begin{equation}
v^{\prime \prime \prime \prime }-(\rho ^{2}-\Gamma -\frac{1}{2}\mathcal{A}%
\left\vert v^{\prime }\right\vert ^{2})v^{\prime \prime }-\gamma ([1-\xi
]v^{\prime })^{\prime }=0,  \label{8}
\end{equation}%
where $v$, as in (\ref{1}) denotes the displacement in the $z$-direction.
The boundary conditions are $v=v^{\prime \prime }=0$ at $\xi =0$ and $\xi =1$%
. Introducing the variable $\alpha ^{2}=\rho ^{2}-(\Gamma +\gamma )$,
integrating once and setting $dv/d\xi =u$ we obtain the equation%
\begin{equation}
u^{\prime \prime }+(\alpha ^{2}+\gamma \xi )u-\frac{1}{2}\mathcal{A}%
\left\vert u^{\prime }\right\vert ^{2}=0,  \label{9}
\end{equation}%
with boundary conditions $u^{\prime }(\xi =0)=u^{\prime }(\xi =1)=0$. The
one-dimensional kernel of the corresponding linear operator 
\begin{equation*}
\emph{L}:=u^{\prime \prime }+(\alpha ^{2}+\gamma \xi )u,
\end{equation*}%
is now spanned not by a sine function, as in the case without gravitational
effects, but by Bessel functions. The proof of this proposition requires
transformation of the equation%
\begin{equation}
u^{\prime \prime }+(\alpha ^{2}+\gamma \xi )u=0,  \label{10}
\end{equation}%
to the equation%
\begin{equation}
t^{2}w^{\prime \prime }+tw^{\prime }+(t^{2}-(\frac{1}{3})^{2})w=0,
\label{11}
\end{equation}%
where in the above equation prime denotes differentiation with respect to $t$%
. First, substitution $(\alpha ^{2}+\gamma \xi )\gamma ^{\sigma }=y,$ $%
\sigma =-2/3,$ leads to the equation%
\begin{equation*}
u^{\prime \prime }+yu=0,
\end{equation*}%
where the prime denotes differentiation with respect to $y.$ The subsequent
substitution $t=\frac{2}{3}y^{3/2},$ $u=\sqrt{y}w$ leads to equation (\ref%
{11}). This equation may be solved in terms of Bessel functions so that the
corresponding solution of the linearized eq. (\ref{8}) is%
\begin{gather}
v(\xi )=(\alpha ^{2}+\gamma \xi )\left[ \text{J}_{2/3}\left( \frac{2}{3}%
\frac{\alpha ^{3}}{\gamma }\right) \text{J}_{-2/3}\left( \frac{2}{3}\frac{%
(\alpha ^{2}+\gamma \xi )^{3/2}}{\gamma }\right) \right.  \notag \\
\left. -\text{J}_{-2/3}\left( \frac{2}{3}\frac{\alpha ^{3}}{\gamma }\right) 
\text{J}_{2/3}\left( \frac{2}{3}\frac{(\alpha ^{2}+\gamma \xi )^{3/2}}{%
\gamma }\right) \right] .  \label{12}
\end{gather}%
For gravity and fluid motion in the same direction, and assuming $\Gamma =0,$
the location of bifurcation points (nontrivial equilibria) is obtained from
the expression%
\begin{equation}
\rho _{n}=n\pi +\frac{1}{4}\frac{\gamma }{n\pi }+\vartheta (\gamma ^{2}),%
\text{ \ \ }n=0,1,2,....  \label{13}
\end{equation}%
Details of the derivation of equations (\ref{10}) and (\ref{13}) are
presented in Appendix B. The above expression suggests that bifurcation
points are shifted as compared to the case without gravity in the direction
of increasing velocity by the amount $\gamma /4n\pi .$ Correspondingly,
bifurcations occur for higher velocities compared to the case when gravity
is not taken into account. As the velocity increases, the effects of gravity
are weaker, so that finally locations of bifurcation points are the same as
for the case without gravity. This is an intuitively pleasing result as one
expects high enough velocity to annul the effects of gravity. If $\ \Gamma $
is not neglected, the expression for critical velocity is%
\begin{gather}
\rho _{n}=n\pi +\frac{1}{4}\frac{\gamma }{n\pi }+\frac{1}{2}\frac{\Gamma }{%
n\pi }+\vartheta (2),  \label{14} \\
=n\pi +\frac{1}{4}\frac{\gamma +2\Gamma }{n\pi }+\vartheta (2),\text{ \ }%
n=1,2,...  \notag
\end{gather}%
The resulting bifurcation diagrams are presented in Fig. 2. Hence the roles
of tensioning and gravity are similar; however, as the velocity increases
the influence of gravity diminishes twice as fast as tensioning effects. The
other aspect of this phenomenon may be observed by considering the distance
between bifurcation points, in units of velocity, given by the following
expression:%
\begin{equation}
\Delta \rho _{n}=\pi -\frac{1}{4\pi }\frac{\gamma }{n(n+1)}-\frac{1}{2\pi }%
\frac{\Gamma }{n(n+1)}.  \label{15}
\end{equation}%
When $\gamma =\Gamma =0,$ bifurcation points are equidistant from each
other, this distance being equal to $\pi .$ When $\gamma $ and $\Gamma $ are
not equal to zero, the distance is an increasing sequence, whose limit $%
(n\rightarrow \infty )$ is equal to $\pi .$

If the direction of the fluid is opposite to the direction of gravity and
assuming no tensioning effects, bifurcation points are located at:%
\begin{equation}
\rho _{n}=n\pi -\frac{1}{4}\frac{\gamma }{n\pi }+\vartheta (\gamma ^{2}),%
\text{ \ \ }n=0,1,2,....  \label{16}
\end{equation}%
In this case bifurcations occur for velocities smaller than in the case
without gravity, and again for high enough velocities the effects of gravity
on the location of bifurcation points are annulled. Inclusion of tensioning
effects yields%
\begin{equation*}
\rho _{n}=n\pi -\frac{1}{4}\frac{\gamma }{n\pi }+\frac{1}{2}\frac{\Gamma }{%
n\pi }+\vartheta (2)=n\pi -\frac{1}{4}\frac{\gamma -2\Gamma }{n\pi }%
+\vartheta (2),\text{ \ \ }n=1,2,...,
\end{equation*}%
and the corresponding bifurcation diagrams are presented in Fig. 3. The
matching bifurcation interpoint distances are 
\begin{equation*}
\Delta \rho _{n}=\pi +\frac{1}{4\pi }\frac{\gamma }{n(n+1)}-\frac{1}{2\pi }%
\frac{\Gamma }{n(n+1)}.
\end{equation*}%
In this case there is a competition between gravity and tensioning effects,
with the possibility that they annul each other when $\gamma =2\Gamma .$

Equation (\ref{11}) yields an approximate solution 
\begin{equation}
v(\xi )\simeq \frac{1}{(\alpha ^{2}+\gamma \xi )^{2}}\sin \left[ \frac{2}{3}%
\frac{(\alpha ^{2}+\gamma \xi )^{3/2}-\alpha ^{3}}{\gamma }\right] ,
\label{17}
\end{equation}%
which for $\gamma \rightarrow 0$ approaches the solution sin$(n\pi \xi )$,
the solution of the linear equation corresponding to (\ref{8}) without
gravitational effects. In order to shed more light on the physical aspect of
the above results, it is instructive to call upon an important relationship
between the solution $\mu (\rho )$ of the algebraic equation $g(\mu ,\rho
)=0 $ and the solution of the full problem $v(\xi )$. Near the bifurcation
point the nontrivial solutions of (\ref{8}) have the spatial structure of
the basis vector $u_{1}$ that spans the one dimensional kernel of the
corresponding linear operator. Hence$,$ the solutions $\mu $ of the reduced
bifurcation equation $g(\mu ,\rho )=0$ are related to solutions of \ Eq.(\ref%
{8}) as%
\begin{equation*}
v=\mu u_{1}+\vartheta (\mu ^{2}),
\end{equation*}%
so that the complete solution of (\ref{8}) in the vicinity of the
bifurcation point may be written as%
\begin{eqnarray}
v(\xi ) &=&\mu (\rho )(\alpha ^{2}+\gamma \xi )\left[ \text{J}_{2/3}\left( 
\frac{2}{3}\frac{\alpha ^{3}}{\gamma }\right) \text{J}_{-2/3}\left( \frac{2}{%
3}\frac{(\alpha ^{2}+\gamma \xi )^{3/2}}{\gamma }\right) \right.  \notag \\
&&\left. -\text{J}_{-2/3}\left( \frac{2}{3}\frac{\alpha ^{3}}{\gamma }%
\right) \text{J}_{2/3}\left( \frac{2}{3}\frac{(\alpha ^{2}+\gamma \xi )^{3/2}%
}{\gamma }\right) +\vartheta (\xi ^{2})\right] ,  \label{18}
\end{eqnarray}%
satisfying boundary conditions $v(0)=v(1)=0;$ $v^{\prime \prime
}(0)=v^{\prime \prime }(1)=0.$

The analysis of the Holmes' model reveals several important features.
Irrespective of the inclusion or exclusion of gravity and/or tensioning
effects, the stationary bifurcation is of supercritical type. Although
gravity and tension have a similar effect on the location of bifurcation
points, only gravity changes the solution locally. In the vicinity of the
bifurcation point, the solution (eigenfunction) is a sine function for $%
\gamma =0$ and a Bessel function for $\gamma \neq 0.$ Moreover, both gravity
and tension increase the distance between bifurcation points along the
parameter space and this effect is particularly noticeable for low
velocities, while high velocities compensate the influence of gravity and
tension as intuitively expected.

\section{The Model of Thurman and Mote}

\subsection{Bifurcations in the stationary model}

The model of Thurman and Mote (Thurman and Mote, 1969; Paidoussis, 1998),
which considers both lateral and axial deflections, was derived under the
following assumptions: (i) no gravity force (ii) steady flow velocity (iii)
linear moment-curvature relationship and (iv) a simple approximation of the
fluid velocity. For the study of the normal form of the bifurcation
equations it is more instructive to consider this model as a special case of
the complete nonlinear model of Paidoussis (2003)\footnote{%
The corrected version of equations is given in Appendix T.4.}, to be
analyzed in the next section. The nondimensional equations for spatial
motion of the complete nonlinear model are%
\begin{gather}
w^{\prime \prime \prime \prime }+(\rho ^{2}-(\Gamma -\Pi ))w^{\prime \prime
}+\gamma w^{\prime }+\left( \Gamma -\mathcal{A}-\Pi \right) (w^{\prime
\prime }u^{\prime }+u^{\prime }w^{\prime \prime }+\frac{3}{2}w^{\prime
2}w^{\prime \prime })  \notag \\
-(3u^{\prime \prime \prime }w^{\prime \prime }+4u^{\prime \prime }w^{\prime
\prime \prime }+2u^{\prime }w^{\prime \prime \prime \prime }+w^{\prime
}u^{\prime \prime \prime \prime }+2w^{\prime \prime 3}+2w^{\prime
2}w^{\prime \prime \prime \prime }+8w^{\prime }w^{\prime \prime }w^{\prime
\prime \prime })  \notag \\
-\gamma \left[ w^{\prime }u^{\prime }+\frac{1}{2}w^{\prime 3}-(1-\xi )\left(
-w^{\prime \prime }+u^{\prime \prime }w^{\prime }+u^{\prime }w^{\prime
\prime }+\frac{3}{2}w^{\prime 2}w^{\prime \prime }\right) \right] =0.
\label{19}
\end{gather}%
\begin{gather}
(\rho ^{2}-\mathcal{A})u^{\prime \prime }-(w^{\prime \prime }w^{\prime
\prime \prime }+w^{\prime }w^{\prime \prime \prime \prime })-\gamma \left[ 
\frac{1}{2}w^{\prime 2}-(1-\xi )w^{\prime }w^{\prime \prime }\right]  \notag
\\
+\left( \Gamma -\mathcal{A}-\Pi \right) w^{\prime }w^{\prime \prime }=0.
\label{20}
\end{gather}%
where the dimensionless tension and the pressure at the downstream end are%
\begin{equation}
\Gamma =\frac{T(L)L^{2}}{EI},\text{ \ \ }\Pi =\frac{P(L)L^{2}}{EI},
\label{21}
\end{equation}%
respectively, while $u$ denotes the dimensionless longitudinal deflection in
the $x$ direction (direction of gravity), and the other quantities being the
same as in the model of Holmes, with $w$ here replacing $v$. A prime denotes
differentiation with respect to the Lagrangian variable $x$ which may be
used interchangeably with $\xi $. The following assumptions may be
attributed to the model of Thurman and Mote:%
\begin{equation*}
\gamma =0,\text{ \ \ \ \ }\Pi =0\text{ \ \ \ \ }\Gamma =\frac{T_{0}L^{2}}{EI}%
,\text{ \ \ \ \ }\kappa ^{2}=0,
\end{equation*}%
where $T_{0}=$const$,$ denotes externally applied tension and $\kappa ^{2}$
denotes curvature. Consequently the equations of motion of this model are%
\begin{eqnarray}
w^{\prime \prime \prime \prime }+(\rho ^{2}-\Gamma )w^{\prime \prime
}+\left( \Gamma -\mathcal{A}\right) (w^{\prime \prime }u^{\prime }+u^{\prime
}w^{\prime \prime }+\frac{3}{2}w^{\prime 2}w^{\prime \prime }) &=&0,  \notag
\\
(\rho ^{2}-\mathcal{A})u^{\prime \prime }+\left( \Gamma -\mathcal{A}\right)
w^{\prime }w^{\prime \prime } &=&0.  \label{22}
\end{eqnarray}%
The boundary conditions are $u(0)=w(0)=u(\xi =1)=w(\xi =1)=0,$ with the
additional condition $w^{\prime \prime }(0)=w^{\prime \prime }(\xi =1)=0$
for a simply-supported pipe or $w^{\prime }(0)=w^{\prime }(\xi =1)=0$ for
clamped-clamped one.

Fixed points, as in the model of Holmes for $\gamma =0$, are located at
positions 
\begin{equation*}
\rho _{n}=n\pi +\frac{1}{2}\frac{\Gamma }{n\pi }+\vartheta (\Gamma ^{2}),%
\text{ \ \ }n=0,1,2,....
\end{equation*}%
as easily seen from the following argument. The kernel of the linear
operator $L$ corresponding to the system consisting of Eqs. (\ref{19}) and (%
\ref{20}) is obtained by solving the equation%
\begin{equation}
L\left( 
\begin{array}{c}
w \\ 
u%
\end{array}%
\right) =\left( 
\begin{array}{c}
w_{0}^{\prime \prime \prime \prime }+(\rho ^{2}-\Gamma )w_{0}^{\prime \prime
} \\ 
u_{0}^{\prime \prime }%
\end{array}%
\right) =\left( 
\begin{array}{c}
0 \\ 
0%
\end{array}%
\right) ,  \label{23}
\end{equation}%
where%
\begin{equation}
\left( 
\begin{array}{c}
w_{0}(x) \\ 
u_{0}(x)%
\end{array}%
\right) =\sum\nolimits_{n=1}^{\infty }\left( 
\begin{array}{c}
a_{n} \\ 
b_{n}%
\end{array}%
\right) \sin n\pi \xi ,  \label{24}
\end{equation}%
satisfying boundary conditions%
\begin{equation*}
\left( 
\begin{array}{c}
w_{0}(0) \\ 
u_{0}(0)%
\end{array}%
\right) =\left( 
\begin{array}{c}
w_{0}(1) \\ 
u_{0}(1)%
\end{array}%
\right) =\left( 
\begin{array}{c}
0 \\ 
0%
\end{array}%
\right) .
\end{equation*}%
Hence, the position of bifurcation points is readily obtained from equations
(\ref{23}) and (\ref{24}). A straightforward calculation indicates that the
kernel of $L$ is spanned by%
\begin{equation}
\text{Ker}L(\left( 
\begin{array}{c}
w \\ 
u%
\end{array}%
,\rho \right) )=\mathbb{R}\left\{ \left( 
\begin{array}{c}
1 \\ 
0%
\end{array}%
\right) \sin n\pi \xi \right\} .  \label{25}
\end{equation}%
Calculation of terms defined in the Appendix A, leads to the following
bifurcation equation 
\begin{equation}
g(\mu ,\rho )=\frac{3}{8}(n\pi )^{4}\left[ \alpha (\beta -3)\right] \text{ }%
\mu ^{3}-(n\pi )^{3}\mu \rho +\vartheta (x^{3})=0,  \label{26}
\end{equation}%
where%
\begin{equation*}
\alpha =\Gamma -\mathcal{A},\text{ \ \ \ \ \ \ \ \ \ }\beta =\frac{\Gamma -%
\mathcal{A}}{\rho ^{2}-\mathcal{A}}.
\end{equation*}%
In the vicinity of the bifurcating solutions, the solutions of the complete
model have the form%
\begin{equation}
\left( 
\begin{array}{c}
w(\xi ) \\ 
u(\xi )%
\end{array}%
\right) =\mu (\rho )\left( 
\begin{array}{c}
1 \\ 
0%
\end{array}%
\right) \sin k\pi \xi +\vartheta (\mu ^{2}),  \label{27}
\end{equation}%
where $\mu (\rho )$ represents the solution of equation (\ref{26}). Hence,
again the bifurcation is of the pitchfork type; however, there are important
differences between the model of Holmes and the model of Thurman and Mote.
It is immediately evident that bifurcation may be of supercritical pitchfork
type for 
\begin{equation*}
\alpha (\beta -3)>0,
\end{equation*}%
and of subcritical type, shown in Fig. 4, for 
\begin{equation*}
\alpha (\beta -3)<0.
\end{equation*}%
Writing explicitly the expression on the left side of these inequalities as%
\begin{equation*}
\frac{(\Gamma -\mathcal{A})^{2}}{\rho _{n}^{2}-\mathcal{A}}-3(\Gamma -%
\mathcal{A}),
\end{equation*}%
and recalling that\ $\rho _{n}^{2}=\Gamma +(n\pi )^{2},$ implies that two
cases may be analyzed, depending on whether $\ |\alpha |/\rho _{n}^{2}<1$ or 
$\rho _{n}^{2}/|\alpha |$ $<1.$ However, since only the first point of
instability corresponding to $n=1$ is relevant from the physical point of
view\ the former condition is of no practical importance although it is
important for understanding the mathematical aspects of the model.

\textit{Case 1}: $\ |\alpha |/\rho _{n}^{2}<1.$ This situation corresponds
to high fluid velocities (large $n$). In this case, a straightforward
calculation yields the following conditions for the occurrence of
supercritical and subcritical bifurcations:

\textit{Supercritical condition}:%
\begin{equation}
\text{ }\Gamma <\mathcal{A}.  \label{28a}
\end{equation}

\textit{Subcritical condition:}%
\begin{equation}
\text{ }\Gamma >\mathcal{A}.  \label{29}
\end{equation}%
Based on (\ref{3}) the above inequalities may be also expressed as%
\begin{equation*}
T_{0}<EA,
\end{equation*}%
and%
\begin{equation*}
T_{0}>EA.
\end{equation*}

\textit{Case 2}: $\rho _{n}^{2}/|\alpha |$ $<1$. This case is applicable to
low fluid velocities (small $n\ $) and large $\left\vert \Gamma -\mathcal{A}%
\right\vert ,$ so that either a highly flexible pipe is considered or the
effects of tensioning are large. Conditions for the occurrence of
supercritical and subcritical bifurcations are

\textit{Supercritical case}:%
\begin{equation}
\Gamma <\mathcal{A-}\frac{1}{2}(n\pi )^{2}.  \label{30}
\end{equation}

\textit{Subcritical case:}%
\begin{equation}
\Gamma >\mathcal{A-}\frac{1}{2}(n\pi )^{2}.  \label{31}
\end{equation}%
For low fluid velocities, both conditions strongly depend on values of $%
\Gamma $ and $\mathcal{A},$ since $n$ is small so that supercritical
bifurcation is more likely in a pipe of high axial flexibility and low
externally applied tension while subcritical bifurcation is more likely in a
pipe with small axial flexibility and high externally applied tension.

Hence, an important feature of the Thurman and Mote model is dependence of
the bifurcation type (supercritical or subcritical) on the relationship
between elastic characteristics of the pipe and externally applied tension,
with the velocity-dependent term figuring only in inequalities relevant at
low fluid velocities (the case of physical validity). At this point it
should be mentioned that if $\rho ^{2}$ does not appear (Eq. (5.62) of
Paidoussis (1998)) in the equation for $u,$ only one classification
condition is obtained in the form of inequalities (\ref{28a}) and (\ref{29}%
); thus from the point of view of the bifurcation theory this version of the
model assumes high fluid velocities. On the other hand, since the case of
high fluid velocities requires large $n$ and is therefore of no practical
importance, the term $\rho ^{2}$ is necessary for the model to be useful in
bifurcation analysis. Comparison with the model of Holmes, which assumes
just lateral deflections and for which only supercritical bifurcation is
possible, reveals that inclusion of the equation for axial deflections makes
subcritical bifurcation also possible.

\subsection{Thurman and Mote model with curvature}

The inclusion of the curvature term $EI\kappa ^{2}$ in the model of Thurman
and Mote may again be considered as a special, reduced version of the
complete nonlinear model. Retaining simplifications of the Thurman and Mote
model in effect, namely $\gamma =0,$\ $\Pi =0,$ $\Gamma =T_{0}L^{2}/EI,$\ $%
\kappa ^{2}=0,$ $T_{0}=$const., but keeping terms that arise from the
curvature effect, the following dimensionless equations of motion are
obtained: 
\begin{gather}
w^{\prime \prime \prime \prime }+(\rho ^{2}-\Gamma )w^{\prime \prime
}+\left( \Gamma -\mathcal{A}\right) (w^{\prime \prime }u^{\prime }+u^{\prime
}w^{\prime \prime }+\frac{3}{2}w^{\prime 2}w^{\prime \prime })  \label{32} \\
-(3u^{\prime \prime \prime }w^{\prime \prime }+4u^{\prime \prime }w^{\prime
\prime \prime }+2u^{\prime }w^{\prime \prime \prime \prime }+w^{\prime
}u^{\prime \prime \prime \prime }+2w^{\prime \prime 3}+2w^{\prime
2}w^{\prime \prime \prime \prime }+8w^{\prime }w^{\prime \prime }w^{\prime
\prime \prime })=0.  \notag
\end{gather}%
\begin{equation}
(\rho ^{2}-\mathcal{A})u^{\prime \prime }-(w^{\prime \prime }w^{\prime
\prime \prime }+w^{\prime }w^{\prime \prime \prime \prime })+\left( \Gamma -%
\mathcal{A}\right) w^{\prime }w^{\prime \prime }=0.  \label{33}
\end{equation}%
Boundary conditions are the same as in the basic Thurman and Mote model.
Arguments used in previously discussed models lead to, as expected, same
locations of fixed points%
\begin{equation*}
\rho _{n}=n\pi +\frac{1}{2}\frac{\Gamma }{n\pi }+\vartheta (\Gamma ^{2}),%
\text{ \ \ }n=0,1,2,....
\end{equation*}

Evaluation of terms of the bifurcation equation shows that it has the
following form:%
\begin{equation}
g(x,\rho )=\frac{3}{4}(n\pi )^{4}[(7-\beta )(n\pi )^{2}-\frac{1}{2}\alpha
(\beta +3)]x^{3}-(n\pi )^{3}x\rho ,  \label{34}
\end{equation}%
where 
\begin{equation*}
\alpha =\Gamma -\mathcal{A},\text{ \ \ \ \ \ \ \ }\beta =\frac{\Gamma -%
\mathcal{A}+2(n\pi )^{2}}{\rho _{n}^{2}-\mathcal{A}}=1+\frac{(n\pi )^{2}}{%
\alpha +(n\pi )^{2}}.
\end{equation*}%
Hence the bifurcation is of supercritical type if 
\begin{equation*}
(7-\beta )(n\pi )^{2}-\frac{1}{2}\alpha (\beta +3)>0,
\end{equation*}%
and subcritical if%
\begin{equation*}
(7-\beta )(n\pi )^{2}-\frac{1}{2}\alpha (\beta +3)<0.
\end{equation*}%
More insight into these inequalities is gained by considering specific cases
dependent on the fluid velocity.

\textit{Case 1}: $\left\vert \alpha /\rho _{n}^{2}\right\vert <1.$ This case
corresponds to high fluid velocities (large $n$).\ A straightforward
calculation leads to conditions for the occurrence of supercritical and
subcritical bifurcations:

\textit{Supercritical condition}:%
\begin{equation}
\text{ }\Gamma <\mathcal{A}+\frac{10}{3}(n\pi )^{2}.  \label{35}
\end{equation}

\textit{Subcritical condition:}%
\begin{equation}
\text{ }\Gamma >\mathcal{A}+\frac{10}{3}(n\pi )^{2}.  \label{36}
\end{equation}%
A very large and hence a nonphysical value of $n$ would be required to make
the term $(n\pi )^{2}$ dominant in the above inequalities. Since with
increasing fluid velocity $\rho $ the effective stiffness of the pipe
diminishes, the model strongly prefers supercritical bifurcation. However,
since only the first mode is relevant from the physical aspect, the interest
is in the case below.

\textit{Case 2}: $\left\vert (n\pi )^{2}/\alpha \right\vert $ $<1$. This
case may be attributed to low fluid velocities (small $n$), and is therefore
of physical and practical interest. The corresponding conditions are:

\textit{Supercritical condition}:%
\begin{equation}
\text{ }\Gamma <\mathcal{A}+\frac{7}{4}(n\pi )^{2}.  \label{37}
\end{equation}

\textit{Subcritical condition:}%
\begin{equation}
\text{ }\Gamma >\mathcal{A}+\frac{7}{4}(n\pi )^{2}.  \label{38}
\end{equation}%
The inclusion of the curvature term in the model increases the effective
axial flexibility (the term $\mathcal{A}+\frac{7}{4}(n\pi )^{2}$) as
compared to the basic Thurman and Mote model (the term $\mathcal{A-}\frac{1}{%
2}(n\pi )^{2}$), so a slight preference is given to the supercritical
bifurcation. Taking into consideration that $n$ is small, tensioning effects
may be dominant, for example in the case of short pipes with high flexural
rigidity.

\subsection{The effect of elastic support}

In order to investigate the essential features of the added elastic support
which involves distributed springs along the length of the pipe, we use the
least complex model, the model of Thurman and Mote model without gravity
effects. The corresponding equations are 
\begin{equation}
w^{\prime \prime \prime \prime }+(\rho ^{2}-\Gamma )w^{\prime \prime
}+\left( \Gamma -\mathcal{A}\right) (w^{\prime \prime }u^{\prime }+u^{\prime
}w^{\prime \prime }+\frac{3}{2}w^{\prime 2}w^{\prime \prime })+Kw=0,
\label{39}
\end{equation}%
\begin{equation}
(\rho ^{2}-\mathcal{A})u^{\prime \prime }+\left( \Gamma -\mathcal{A}\right)
w^{\prime }w^{\prime \prime }=0.  \label{40}
\end{equation}%
The algebraic bifurcation equation is identical to eq. (\ref{26}), hence
elastic support does not change the form of the solution in the vicinity of
bifurcation points. From the equation for the kernel of the corresponding
linear operator 
\begin{equation*}
L\left( 
\begin{array}{c}
w \\ 
u%
\end{array}%
\right) =\left( 
\begin{array}{c}
w^{\prime \prime \prime \prime }+(\rho ^{2}-\Gamma )w^{\prime \prime }+Kw \\ 
u^{\prime \prime }%
\end{array}%
\right) =\left( 
\begin{array}{c}
0 \\ 
0%
\end{array}%
\right) ,
\end{equation*}%
and eq. (\ref{24}) the relation determining position of bifurcation points
is obtained:%
\begin{gather}
\rho _{n}=(n\pi )\left[ 1+\frac{\Gamma }{(n\pi )^{2}}+\frac{K}{(n\pi )^{4}}%
\right] ^{1/2}  \notag \\
=n\pi +\frac{1}{2}\frac{\Gamma }{(n\pi )}+\frac{1}{2}\frac{K}{(n\pi )^{3}}%
+\vartheta (2),\text{ \ \ \ }n=0,1,2,....  \label{41}
\end{gather}

Clearly if the spring constant is small, then the nontrivial bifurcation
points are located at same positions $\rho _{n}$ as in the Thurman and Mote
model. The effect or elastic support is important for very small $n$ (e.g. $%
n=1$ or $n=2$) and large values of $K$. However, if $K$ is large, and $%
(1/2)\Gamma /n\pi $ small in comparison with $(n\pi ),$ their locations may
be distributed as in Fig. 5. A brief analysis of the above expression shows
that the first nontrivial bifurcation point and some of the subsequent ones
(usually for the first few $n^{\prime }$s ) are to a large degree determined
by the value of$\ K,$ and the same applies to the distances between the
bifurcation points. Afterwards the term $(n\pi )$ dominates and the
bifurcation points are equally distributed. A complete nonlinear model
consisting of equations (\ref{19}) and (\ref{20}) and including the term
corresponding to the elastic support is very interesting both from
mathematical and physical aspects; however, due to its complexity it will be
analyzed elsewhere (Rajkovi\'{c}, 2005).

\section{The complete nonlinear model of Paidoussis}

\subsection{Symmetry considerations}

The derivation of the complete nonlinear model, based on the work (Semler et
al., 1994), in its correct form is given in Paidoussis (2003). This model,
as presented in Section 2, in contrast to the previously analyzed models,
includes effects of gravity and pressure at the downstream end, so the
dimensionless equations of motion of an extensible cylinder conveying fluid
have the form:

\begin{gather}
w^{\prime \prime \prime \prime }+(\rho ^{2}-(\Gamma -\Pi ))w^{\prime \prime
}+\gamma w^{\prime }+\left( \Gamma -\mathcal{A}-\Pi \right) (w^{\prime
\prime }u^{\prime }+u^{\prime }w^{\prime \prime }+\frac{3}{2}w^{\prime
2}w^{\prime \prime })  \notag \\
-(3u^{\prime \prime \prime }w^{\prime \prime }+4u^{\prime \prime }w^{\prime
\prime \prime }+2u^{\prime }w^{\prime \prime \prime \prime }+w^{\prime
}u^{\prime \prime \prime \prime }+2w^{\prime \prime 3}+2w^{\prime
2}w^{\prime \prime \prime \prime }+8w^{\prime }w^{\prime \prime }w^{\prime
\prime \prime })  \notag \\
-\gamma \left[ w^{\prime }u^{\prime }+\frac{1}{2}w^{\prime 3}-(1-\xi )\left(
-w^{\prime \prime }+u^{\prime \prime }w^{\prime }+u^{\prime }w^{\prime
\prime }+\frac{3}{2}w^{\prime 2}w^{\prime \prime }\right) \right] =0,
\label{42}
\end{gather}%
\begin{gather}
(\rho ^{2}-\mathcal{A})u^{\prime \prime }-(w^{\prime \prime }w^{\prime
\prime \prime }+w^{\prime }w^{\prime \prime \prime \prime })-\gamma \left[ 
\frac{1}{2}w^{\prime 2}-(1-\xi )w^{\prime }w^{\prime \prime }\right]  \notag
\\
+\left( \Gamma -\mathcal{A}-\Pi \right) w^{\prime }w^{\prime \prime }=0.
\label{43}
\end{gather}%
The relevant quantities have been defined in the previous section.

To begin with, it is of interest to inspect the symmetry properties of this
equation. It is immediately clear that in the first equation, only odd
powers of $w$ turn up, while this is not the case with the second equation.\
Hence, solution of the set (\ref{42}) and (\ref{43}), $v=(w$ $\ u)^{\text{T}%
},$ satisfies the following symmetry condition%
\begin{equation}
T\left( 
\begin{array}{c}
w(\xi ) \\ 
u(\xi )%
\end{array}%
\right) =\left( 
\begin{array}{c}
-w(\xi ) \\ 
u(\xi )%
\end{array}%
\right) ,  \label{44}
\end{equation}%
where $T$ represents the operator of the symmetry group. In the vicinity of
bifurcating solutions, the solutions of the complete model have the form%
\begin{equation}
\left( 
\begin{array}{c}
w(\xi ) \\ 
u(\xi )%
\end{array}%
\right) =\mu (\rho )\left( 
\begin{array}{c}
w_{0}(\xi ) \\ 
0%
\end{array}%
\right) +\vartheta (\mu ^{2}),  \label{45}
\end{equation}%
where $w_{0}(x)$ represents the solution of the linear equation 
\begin{equation}
w^{\prime \prime \prime \prime }+[(\rho ^{2}-(\Gamma -\Pi ))-\gamma (1-\xi
)]w^{\prime \prime }+\gamma w^{\prime }=0.  \label{46}
\end{equation}%
Acting with the symmetry operator $T$ on equation (\ref{45}) one obtains%
\begin{gather*}
T\left( 
\begin{array}{c}
w(\xi ) \\ 
u(\xi )%
\end{array}%
\right) =\mu (\rho )\left( 
\begin{array}{c}
-w_{0}(\xi ) \\ 
0%
\end{array}%
\right) +\vartheta (\mu ^{2}), \\
=-\mu (\rho )\left( 
\begin{array}{c}
w_{0}(\xi ) \\ 
0%
\end{array}%
\right) +\vartheta (\mu ^{2}).
\end{gather*}%
Hence, the bifurcation equation satisfies the relationship%
\begin{equation*}
g(-\mu ,\rho )=-g(\mu ,\rho )
\end{equation*}%
so that $g$ possesses the \textbf{Z}$_{2}$ symmetry. The above symmetry
properties indicate that the bifurcation is necessarily of the pitchfork
type (Golubitsky, 1985). The physical representation of this symmetry is a
reflection across the longitudinal pipe axis. The use of only one equation
in the model of Holmes, the one involving $w,$ may be justified based on
these symmetry considerations.

\subsection{Critical velocity}

With the fluid velocity as the bifurcation parameter, the procedure for
obtaining critical velocity values corresponding to bifurcation points is
almost the same as in the model of Holmes. Specifically, equation (\ref{46}) 
\begin{gather}
w^{\prime \prime \prime \prime }+[(\rho ^{2}-\Gamma +\Pi )-\gamma (1-\xi
)]w^{\prime \prime }+\gamma w^{\prime }  \notag \\
=w^{\prime \prime \prime \prime }+(\rho ^{2}-\Gamma +\Pi )w^{\prime \prime
}-\gamma \frac{d}{d\xi }[(1-\xi )w^{\prime }]=0.  \label{47}
\end{gather}%
corresponds to the linearized version of equation (\ref{8}) of the Holmes'
model. The only difference is the inclusion of the pressure term. The
equation analogous to equation (B.8) in Appendix B determining bifurcation
points is:%
\begin{equation*}
\frac{2}{3}\frac{(\zeta ^{2}+\gamma )^{3/2}-\zeta ^{3}}{\gamma }=n\pi ,
\end{equation*}%
where 
\begin{equation}
\zeta ^{2}=\rho ^{2}-(\Gamma -\Pi )-\gamma .  \label{48}
\end{equation}%
Taylor expanding and retaining terms to second order yields the following
expression for the critical velocity values (bifurcation points):

\begin{gather}
\rho _{n}(L)=n\pi +\frac{1}{4}\frac{\gamma }{n\pi }+\frac{1}{2}\frac{(\Gamma
-\Pi )}{n\pi }+\vartheta (2)  \notag \\
=n\pi +\frac{1}{4}\frac{\gamma +2(\Gamma -\Pi )}{n\pi }+\vartheta (2),\text{
\ }n=1,2,...  \label{49}
\end{gather}%
An important feature of the critical velocity $\rho _{n},$ in contrast to
previously considered less complex models, is its dependence on the pipe
length through the length dependence of $\Gamma $ and $\ \Pi .$ This
dependence clearly diminishes with increasing velocity. Expression (\ref{49}%
) is obtained assuming that fluid flow is in the direction of gravity $%
(\gamma >0).$ Bifurcations occur for velocity values higher than in the case
when gravity, pressure and tension are not taken into account, under the
assumption that%
\begin{equation}
\gamma +2\Gamma >2\Pi .  \label{50}
\end{equation}%
The influence of gravity and tension is diminished by the pressure at the
downstream end as intuitively expected. The individual effect of the gravity
term becomes dominant if the pressure value approaches the value of tension.
If the flow is in the direction opposite to the direction of gravity $%
(\gamma <0),$ the expression for bifurcation points is%
\begin{gather}
\rho _{n}(L)=n\pi -\frac{1}{4}\frac{\gamma }{n\pi }+\frac{1}{2}\frac{(\Gamma
-\Pi )}{n\pi }+\vartheta (2)  \notag \\
=n\pi -\frac{1}{4}\frac{\gamma -2(\Gamma -\Pi )}{n\pi }+\vartheta (2),\text{
\ }n=1,2,...  \label{51}
\end{gather}%
Clearly, gravity in this case acts in the same direction as pressure, and
together they oppose the effects of tension. Additional insight into the
position of bifurcation points, in units of velocity, may be obtained by
considering expressions for interpoint distances. For $\gamma >0$ the
interpoint distance is%
\begin{gather}
\Delta \rho _{n}(L)=\pi -\frac{1}{4\pi }\frac{\gamma }{n(n+1)}-\frac{1}{2\pi 
}\frac{\Gamma -\Pi }{n(n+1)}  \notag \\
=\pi -\frac{1}{4\pi n(n+1)}[\gamma -2(\Gamma -\Pi )].  \label{52}
\end{gather}%
As evident from the above expression, at low fluid velocities this distance
is a sequence of increasing values provided that%
\begin{equation*}
\gamma +2\Pi >2\Gamma ,
\end{equation*}%
and a sequence of decreasing values for%
\begin{equation*}
\gamma +2\Pi <2\Gamma .
\end{equation*}%
Since $\gamma $ is length dependent [Eq. (\ref{2})], the former condition is
more easily satisfied for long pipes, while the latter is more likely
fulfilled for short pipes. For large $n$ (high velocities), distance between
bifurcation points has a fixed value of $\pi .$ For $\gamma <0,$ expression (%
\ref{52}) becomes%
\begin{gather*}
\Delta \rho _{n}(L)=\pi +\frac{1}{4\pi }\frac{\gamma }{n(n+1)}-\frac{1}{2\pi 
}\frac{\Gamma -\Pi }{n(n+1)} \\
=\pi +\frac{1}{4\pi n(n+1)}[\gamma -2(\Gamma -\Pi )].
\end{gather*}%
In this case the distances between the bifurcation points form a decreasing
sequence if 
\begin{equation*}
\gamma +2\Pi >2\Gamma ,
\end{equation*}%
and an increasing sequence provided that%
\begin{equation*}
\gamma +2\Pi <2\Gamma .
\end{equation*}%
Recalling that $\gamma <0,$ the first condition is more likely in short
pipes while the second one is more probable in long pipes.

\subsection{Normal form of the bifurcation equation}

The normal form of the pitchfork bifurcation modulo higher order terms
(hot), reads 
\begin{equation}
g(\mu ,\rho )=g_{\mu \mu \mu }\mu ^{3}+g_{\mu \rho }\mu \rho +\text{hot.}
\label{53}
\end{equation}%
In Appendix B, terms $g_{\mu \mu \mu }$ and $g_{\mu \rho }$ are defined with
the symbol $x$ replacing $\mu .$ In previously considered models the term $%
g_{\mu \rho }$ was explicitly determined and in all cases it was negative.
Hence, the sign of $g_{\mu \mu \mu }$ determines whether bifurcation is of
supercritical or subcritical type. For the complete nonlinear model,
evaluation of terms $g_{\mu \mu \mu }$ and $g_{\mu \rho }$ is much more
complicated, and in order to minimize computational effort it is sufficient
to determine just the sign of $g_{\mu \rho }$. Following evaluation of the
expression for $g_{\mu \rho }$ given in the Appendix A, the following
relationship is obtained:%
\begin{equation}
g_{\mu \rho }=\left\langle v_{0}^{\ast },L_{\rho }\cdot v_{0}\right\rangle
=2\rho \left\langle v_{0}^{\ast },v_{0}^{\prime \prime }\right\rangle ,
\label{54}
\end{equation}%
where $L$ is the linear operator corresponding to the set of equations (\ref%
{42}) and (\ref{43}) 
\begin{equation}
L=\left( 
\begin{array}{c}
w_{0}^{\prime \prime \prime \prime }+[(\rho ^{2}-(\Gamma -\Pi ))-\gamma
(1-\xi )]w_{0}^{\prime \prime }+\gamma w_{0}^{\prime } \\ 
u_{0}^{\prime \prime }%
\end{array}%
\right) .  \label{55}
\end{equation}%
$v_{0}$ represents the solution of $L=0$ evaluated using procedure presented
in Appendix B. With $\zeta $ given by expression (\ref{48}) $v_{0}$ has the
following form%
\begin{eqnarray}
v_{0} &=&\left( 
\begin{array}{c}
w_{0}(\xi ) \\ 
u_{0}(\xi )%
\end{array}%
\right) =  \notag \\
&&\left( 
\begin{array}{c}
1 \\ 
0%
\end{array}%
\right) (\zeta ^{2}+\gamma \xi )\left[ \text{J}_{2/3}\left( \frac{2}{3}\frac{%
\zeta ^{3}}{\gamma }\right) \text{J}_{-2/3}\left( \frac{2}{3}\frac{(\zeta
^{2}+\gamma \xi )^{3/2}}{\gamma }\right) \right.  \notag \\
&&\left. -\text{J}_{-2/3}\left( \frac{2}{3}\frac{\zeta ^{3}}{\gamma }\right) 
\text{J}_{2/3}\left( \frac{2}{3}\frac{(\zeta ^{2}+\gamma \xi )^{3/2}}{\gamma 
}\right) \right] ,  \label{56}
\end{eqnarray}%
satisfying boundary conditions%
\begin{equation*}
v_{0}(0)=v_{0}(1)=0.
\end{equation*}%
Explicit expression of the inner product in (\ref{54}) yields%
\begin{equation*}
g_{\mu \rho }=2\rho \dint\nolimits_{0}^{1}v_{0}v_{0}^{\prime \prime }d\xi .
\end{equation*}%
Performing integration by parts and using boundary conditions, one obtains%
\begin{equation*}
g_{\mu \rho }=2\rho \lbrack (v_{0}v_{0}^{\prime
})|_{0}^{1}-\dint\nolimits_{0}^{1}v_{0}^{2}d\xi =-2\rho
\dint\nolimits_{0}^{1}v_{0}^{2}d\xi .
\end{equation*}%
Since the velocity $\rho >0$ and 
\begin{equation*}
\dint\nolimits_{0}^{1}v_{0}^{2}d\xi >0,
\end{equation*}%
it follows that%
\begin{equation}
g_{\mu \rho }=-2\rho \dint\nolimits_{0}^{1}v_{0}^{2}d\xi <0.  \label{57}
\end{equation}%
Hence, once the sign of $g_{\mu \rho }$ is known an explicit evaluation of
this term is unnecessary. However, an explicit evaluation of the term $%
g_{\mu \mu \mu }$ is essential in order to obtain conditions determining the
bifurcation type.

Details related to the evaluation of $g_{\mu \mu \mu }$, due to its
mathematical complexity and extensiveness, are presented in Appendix C. Only
results relevant for the final form of the expression determining the sign
of $g_{\mu \mu \mu }$ are presented in this section. Following appropriate
calculations $g_{\mu \mu \mu }$ assumes the following form:%
\begin{gather}
g_{\mu \mu \mu }=-\left[ \Lambda (\zeta )\frac{(\Gamma -\mathcal{A}-\Pi
+\gamma )}{\rho _{n}^{2}-\mathcal{A}}\left( \frac{2}{3}\frac{\zeta ^{3}}{%
\gamma }\right) \right.  \notag \\
\left. +\Delta (\zeta )\frac{1}{\rho _{n}^{2}-\mathcal{A}}\left( \frac{2}{%
3\gamma }\right) ^{1/3}\zeta +\Omega (\zeta )\right] .  \label{58}
\end{gather}%
Terms $\Lambda (\zeta )$, $\Delta (\zeta )$ and $\Omega (\zeta )$ are
polynomial functions of $\zeta $ and their explicit representation may be
found in Appendix C. As in previous, less complicated models, two cases may
be considered.

\textit{Case 1:} $\gamma /\zeta ^{2}<1.$ In the high velocity limit $\rho
_{n}^{2}-\mathcal{A}$ $>0,$ so that the dominant part of (\ref{58})
determining the sign of $g_{\mu \mu \mu }$ is 
\begin{equation}
-2\Lambda (\zeta )(\Gamma -\mathcal{A}-\Pi +\gamma ).  \label{59}
\end{equation}%
Hence, the bifurcation is supercritical or subcritical if the above
expression is greater then or less then zero respectively. In order to be
precise, it is informative to consider expression $\Lambda (\zeta )$.
Explicitly, this expression is%
\begin{gather}
\Lambda (\zeta )=\left[ a\left( \frac{2}{3}\gamma \right) ^{2}\left( \text{J}%
_{-2/3}\left( \frac{2}{3}\frac{\zeta ^{3}}{\gamma }\right) \right) ^{3}\text{%
J}_{2/3}\left( \frac{2}{3}\frac{\zeta ^{3}}{\gamma }\right) \right.  \notag
\\
\left. +b\left( \frac{2}{3}\gamma \right) ^{8/3}\left( \text{J}_{-2/3}\left( 
\frac{2}{3}\frac{\zeta ^{3}}{\gamma }\right) \right) ^{4}\right] ,
\label{60}
\end{gather}%
where $a$ and $b$ are positive constants:

\begin{gather*}
a=\frac{1}{\left( \Gamma \left( \frac{2}{3}\right) \right) ^{2}}\left( \frac{%
1}{\Gamma \left( \frac{1}{3}\right) }+\frac{1}{\Gamma \left( \frac{2}{3}%
\right) }\right) ^{2} \\
b=a\left( \frac{1}{\Gamma \left( \frac{1}{3}\right) }+\frac{1}{\Gamma \left( 
\frac{2}{3}\right) }\right) ^{-1},
\end{gather*}%
and $\Gamma \left( .\right) $ is the gamma function. The sign of $\Lambda
(\zeta )$ is also positive as 
\begin{equation*}
b\left( \frac{2}{3}\gamma \right) ^{8/3}\left( \text{J}_{-2/3}\left( \frac{2%
}{3}\frac{\zeta ^{3}}{\gamma }\right) \right) ^{4}>a\left( \frac{2}{3}\gamma
\right) ^{2}\left( \text{J}_{-2/3}\left( \frac{2}{3}\frac{\zeta ^{3}}{\gamma 
}\right) \right) ^{3}\text{J}_{2/3}\left( \frac{2}{3}\frac{\zeta ^{3}}{%
\gamma }\right) ,
\end{equation*}%
irrespective of the possible negative signs of either J$_{2/3}$ or J$%
_{-2/3}. $ Therefore, conditions for supercritical and subcritical
bifurcations are determined by the term $-(\Gamma -\mathcal{A}-\Pi +\gamma
), $ so that conditions are:

\textit{Supercritical condition}:%
\begin{equation}
\Gamma +\gamma <\mathcal{A}+\Pi ,  \label{61}
\end{equation}

\textit{Subcritical condition:}%
\begin{equation}
\Gamma +\gamma >\mathcal{A}+\Pi .  \label{62}
\end{equation}%
Recalling expressions for dimensionless system parameters (\ref{2}), (\ref{3}%
) and (\ref{21}), inequalities (\ref{61}) and (\ref{62}) may be also
expressed as:%
\begin{equation}
T(L)+(M+m)gL<EA+P(L),  \label{63}
\end{equation}%
and%
\begin{equation}
T(L)+(M+m)gL>EA+P(L).  \label{64}
\end{equation}%
revealing their dependence on the length of the pipe. Physically, the
effects of gravity are related to the length of the pipe [equation (\ref{2}%
)], so larger $\gamma $ may be associated with a longer pipe and smaller $%
\gamma $ with a shorter one. For short metal pipes $\gamma $ is rather small
and its effects in inequalities (\ref{61}) and (\ref{62}) are weak. A
comparison with corresponding inequalities for the Thurman and Mote model
which includes curvature effects, shows that the pressure term replaces the
velocity-dependent term. The reason is that the pressure term acts the same
as the velocity term and it is clear that an adequately high level of
pressurization alone may cause supercritical bifurcation. For pipes made of
elastic material, gravity effects are important, as are the effects of axial
flexibility, so that a supercritical bifurcation is more probable in shorter
pipes. Intuitively, short pipes acquire a buckled shape in an evolutionary
manner (corresponding to the supercritical case), while longer pipes
suddenly deform (corresponding to the subcritical case). For long pipes
tensioning term is small (internal dissipation of the pipe material is
assumed to be of the Kelvin-Voigt type) so that subcritical bifurcation is
more likely to occur for low levels of pressurization. Thus, high
pressurization in short pipes makes supercritical bifurcation more probable,
while increasing the length of the pipe enhances the probability of
subcritical bifurcation.

\textit{Case 2:} $\zeta ^{2}/\gamma <1.$ Assuming that $\rho _{n}^{2}-%
\mathcal{A}$ $<0$ in this low velocity case (which is of real physical
importance) the sign-dominant term is%
\begin{equation}
\varphi +(\Gamma -\mathcal{A}-\Pi +\gamma )\psi ,  \label{65}
\end{equation}%
where $\varphi $ and $\psi $ are the following expressions:%
\begin{align}
\varphi & =-c_{1}\left( \frac{2}{3}\gamma \right) ^{10/3}\left( \text{J}%
_{-2/3}\left( \frac{2}{3}\frac{\zeta ^{3}}{\gamma }\right) \right)
^{2}\left( \text{J}\left( \frac{2}{3}\frac{\zeta ^{3}}{\gamma }\right)
\right) ^{2}  \label{66} \\
& -c_{2}\left( \frac{2}{3}\gamma \right) ^{8/3}\left( \text{J}_{-2/3}\left( 
\frac{2}{3}\frac{\zeta ^{3}}{\gamma }\right) \right) ^{4}  \notag \\
& -c_{3}\left( \frac{2}{3}\gamma \right) ^{7/3}\left( \text{J}_{-2/3}\left( 
\frac{2}{3}\frac{\zeta ^{3}}{\gamma }\right) \right) ^{3}\text{J}%
_{2/3}\left( \frac{2}{3}\frac{\zeta ^{3}}{\gamma }\right) ,  \notag \\
\psi & =\left( \text{J}_{-2/3}\left( \frac{2}{3}\frac{\zeta ^{3}}{\gamma }%
\right) \right) ^{4}\left[ \left( \frac{2}{3}\gamma \right) ^{7/3}+\left( 
\frac{2}{3}\gamma \right) ^{2}\frac{2^{1/3}}{\left( \Gamma \left( \frac{2}{3}%
\right) \right) ^{2}}\right] \left( \frac{1}{\Gamma \left( \frac{1}{3}%
\right) }+\frac{1}{\Gamma \left( \frac{2}{3}\right) }\right) ,  \notag
\end{align}%
and where $c_{1},$ $c_{2}$ and $c_{3}$ are constants (given in Appendix C).
Clearly $\psi >0$ and it is easy to notice that the sign of $\varphi <0$
does not depend on signs of J$_{-2/3}\left( (2/3)\zeta ^{3}/\gamma \right) $
and J$_{2/3}\left( (2/3)\zeta ^{3}/\gamma \right) ,$ since the term
containing these two Bessel functions is much smaller then the other two
terms. Hence, the following conditions are obtained for bifurcation types:

\textit{Supercritical condition}:%
\begin{equation}
\Gamma +\gamma >\mathcal{A}+\Pi +\frac{\left\vert \varphi \right\vert }{\psi 
},  \label{67}
\end{equation}

\textit{Subcritical condition:}%
\begin{equation}
\Gamma +\gamma <\mathcal{A}+\Pi +\frac{\left\vert \varphi \right\vert }{\psi 
}.  \label{68}
\end{equation}%
Compared to conditions (\ref{61}) and (\ref{62}), inequalities pertaining to
the low velocity case have an additional term acting in the same manner as
the pressure and axial flexibility. The term $\left\vert \varphi \right\vert
/\psi $ is approximately proportional to $\gamma $ so it counterbalances the
gravity term on the left side of inequalities and the above inequalities may
be further replaced with expressions%
\begin{equation*}
T(L)>EA+P(L),
\end{equation*}%
and 
\begin{equation*}
T(L)<EA+P(L),
\end{equation*}%
for subcritical and supercritical cases respectively. Since tension is
inversely proportional to the length of the pipe, for long pipes subcritical
condition is practically always satisfied. For short pipes supercritical
bifurcation is possible only if pressurization term is small compared to
axial flexibility, since the corresponding condition may be written as%
\begin{equation}
EA\left( \frac{1}{2L}\dint\nolimits_{0}^{L}w^{\prime 2}d\xi -1\right) >P(L).
\label{69}
\end{equation}%
In the above expression a viscoelastic material has been considered, and
since we are considering a time independent model, axial tension is 
\begin{equation*}
T=\sigma A=(E\varepsilon +E^{\ast }\dot{\varepsilon})A=\varepsilon EA.
\end{equation*}%
where averaged axial strain $\varepsilon $ due to lateral deflections $w$ is 
\begin{equation*}
\varepsilon =\frac{1}{2L}\dint\nolimits_{0}^{L}w^{\prime 2}d\xi .
\end{equation*}%
The other possibility for this low velocity case, $\rho _{n}^{2}-\mathcal{A}$
$<0,$ has no physical meaning since $n$ is small.

Although a subtle interplay between system parameters requires careful
analysis under specific circumstances of interest, in general it may be
concluded that in the case of high fluid velocity a supercritical
bifurcation is more likely in a short pipe, while a subcritical bifurcation
is more probable in a long pipe (provided that appropriate inequalities (\ref%
{61}) and (\ref{62}) are fulfilled). In the case of low fluid velocities
(along with $\rho _{n}^{2}>\mathcal{A)}$ the situation is the opposite: a
supercritical bifurcation is more likely in a long pipe, while subcritical
bifurcation is preferred in short pipes provided condition (\ref{69} ) is
satisfied.

\section{Conclusion}

Exact analytical solutions in the vicinity of bifurcation point for each
model are obtained along with the derivation of conditions that classify
bifurcations as supercritical or subcritical. Moreover, the analysis is
performed in such a manner that influence of important quantities on the
dynamics of supported fluid conveying pipes may be analyzed in the light of
increasing complexity of each model.

Two important features of the stationary bifurcations for the supported
fluid-conveying pipes should be emphasized. First, all bifurcations are of
the pitchfork type as a consequence of reflection symmetry. Second, all
perturbations of the pitchfork bifurcation, due to gravity or curvature
effects for example, preserve the topological form of the unperturbed
bifurcating diagram, due to the fact that 0 remains the solution of the
perturbed equation. Consequently, unfolding of the bifurcation does not take
place which requires that 0 is not the solution of the perturbed equation.

Models that consider both axial and lateral deflections, hence two-equation
models, enable the possibility of both supercritical and subcritical
pitchfork bifurcations. In contrast, the single equation model of Holmes
which considers just transverse deflections allows only supercritical
bifurcations. An important general characteristic of classification of
generic codimension-1 bifurcations into supercritical or subcritical is that
the bifurcation type depends on only two factors: nonlinear terms and
boundary conditions. Hence, nonlinear terms figuring in the equation for
axial deflections make an important contribution to the terms defining the
normal form of the bifurcation equation.

In the complete nonlinear model of Paidoussis critical velocity values at
which bifurcation occurs depend on tension, gravity and pressure. For $%
\gamma >0$ gravity and tensioning exert matching effects reflected in
shifting bifurcation values in the positive direction with the shift due to
gravity being one-half the corresponding tensioning shift. The effects of
pressurization oppose effects of gravity and tension. For $\gamma <0,$
gravity and pressurization act in the same direction, while tensioning
exerts opposing effect. The critical velocity values for each model are
presented in Table 1. A summary of conditions classifying bifurcations as
supercritical and subcritical is presented in Tables 2 and 3. An increase of
complexity of the bifurcation type conditions may be traced, starting with
the least complex model of Thurman and Mote and ending with the complete
nonlinear model of Paidoussis. The velocity-dependent term appearing in
inequalities corresponding to the Thurman and Mote model with curvature
transforms into an analogous pressurizing effect in the most complex model.
Setting $\gamma =\Pi =0,$ and assuming that $\Gamma $ and $\Pi $ are
independent of the pipe length in the complete nonlinear model of
Paidoussis, one obtains classification conditions for the Thurman and Mote
model. Although conditions for determining whether bifurcation is of
supercritical or subcritical type involve a delicate interaction among
system parameters (gravity effects, tensioning, pressurization and axial
flexibility), it may be concluded that a general tendency is that for high
fluid velocities supercritical bifurcation is more likely in short pipes,
while subcritical is more likely in longer pipes (provided that appropriate
inequalities (\ref{61}) and (\ref{62}) are satisfied). When the fluid
velocity is low,\ but still high enough that the square of it exceeds axial
flexibility, a supercritical bifurcation is more likely in longer pipes
(with conditions (\ref{67}) and (\ref{68}) in effect), while a subcritical
bifurcation may occur in short pipes provided that condition (\ref{69}) is
satisfied.

\section{Acknowledgement}

The authors wish to thank anonymous referees for helpful and instructive
comments, suggestions and criticisms. This work is partially supported by
the Serbian Ministry of Science and Technology as part of the project OI
1986.\bigskip

\newpage \noindent\textbf{Appendix A. \ Lyapunov-Schmidt reduction}

Reduction of a nonlinear equation or a system of nonlinear equations 
\begin{equation}
\Phi _{i}(y,\Lambda )=0,\text{ \ \ \ \ }i=1,...,n  \tag{A.1}  \label{a.1}
\end{equation}%
to the single algebraic equation $g(x,\lambda )$ is the essential feature of
the Lyapunov-Schmidt procedure. The vector $y=y(y_{1},...,y_{n})$ is the
unknown in the above equation, while $\Lambda $ is a vector of parameters.
We assume that only one parameter $\lambda $ is of concern. The main
starting assumption is%
\begin{equation}
\Phi _{i}(0,0)=0,  \tag{A.2}  \label{a.2}
\end{equation}%
and of interest is to describe the solutions of this system locally, in the
vicinity of the origin. If the rank of the $\ n$ $\times $ $n$ Jacobian
matrix $L=(d\Phi )_{0,0},$ is equal to $n$ (nondegenerate case), the
implicit function theorem guarantees the existence of solution $y$ as a
function of $\lambda .$ If the rank is not equal to the size of the Jacobian
matrix, we assume the minimally degenerate case, i.e. 
\begin{equation*}
\text{rank }L=n-1.
\end{equation*}%
Assuming that $\Phi :\mathbb{R}^{n}\times \mathbb{R}^{k}\rightarrow \mathbb{R%
}^{n}$ is a smooth mapping, vector space complements $M$ and $N$ are chosen
to $\ker $ $L$ and range $L$, respectively, so that%
\begin{equation*}
\mathbb{R}^{n}=\ker L\oplus \text{ }M,
\end{equation*}%
and%
\begin{equation*}
\mathbb{R}^{n}=N\oplus \text{range }L,
\end{equation*}%
where $N$ denotes the null space of $L$. Introducing the projection operator 
$E:\mathbb{R}^{n}\rightarrow $ range $L$, the starting system of equation is
expanded into%
\begin{gather}
E\Phi (y,\lambda )=0,  \tag{A.3}  \label{a.3} \\
(I-E)\Phi (y,\lambda )=0,  \notag
\end{gather}%
where $I-$ $E$ is the complementary projection operator to $E.$ Solving the
first equation of (\ref{a.3}) which fulfills the conditions of the implicit
function theorem for $n-1$ of the $\ y$ variables, and inserting solutions
in the second equation, yields an equation for the remaining one variable.
Because of the splitting of $\mathbb{R}^{n},$ any vector $y\in \mathbb{R}%
^{n} $ may written as $y=v(\in \ker L$) $+$ $w(\in M)$, so that mapping%
\begin{equation*}
F:(\ker \text{ }L)\times M\times \mathbb{R}^{k}\rightarrow \text{range }L,
\end{equation*}%
is given by expression%
\begin{equation*}
F(v,w,\lambda )=E\Phi (v+w,\lambda ).
\end{equation*}%
The linear map%
\begin{equation*}
L:M\rightarrow \text{range }L
\end{equation*}%
is invertible, thus according to the implicit function theorem it may be
solved for $w$ near the origin. Denoting this solution as $w=W(v,\lambda
):\ker L\times $ $\mathbb{R}^{k}\rightarrow M,$ which satisfies%
\begin{equation*}
E\Phi (v+W(v,\lambda ),\lambda )=0,\text{ \ \ }W(0,0)=0,
\end{equation*}%
and it may be substituted into the second equation of (\ref{a.3}) to obtain
the reduced mapping $\phi :\ker $ $L\times \mathbb{R}^{k}\rightarrow N,$
where 
\begin{equation*}
\phi (v,\lambda )=(I-E)\Phi (v+W(v,\lambda ),\lambda ).
\end{equation*}%
Consequently, the zeros of the $\phi (v,\lambda )$ are in one-to-one
correspondence with the zeros of $\Phi (y,\lambda ),$ or explicitly%
\begin{equation*}
\phi (v,\lambda )=0\text{ \ if and only if }\Phi (v+W(v,\lambda ),\lambda
)=0.
\end{equation*}%
The reduced function $\phi (v,\lambda )$ may be further used to obtain the
algebraic equation%
\begin{equation*}
g(x,\lambda )=\left\langle v_{0}^{\ast },\phi (xv_{0},\lambda )\right\rangle
,
\end{equation*}%
where $v_{0}^{\ast }\in (range$ $L)^{\perp },$ and $\langle .,.\rangle $
represents the standard inner product. Illustration of the reduction
procedure is shown in Fig. A1. Finally, the derivatives of $g$ figuring in
the Taylor expansion of $g(x,\lambda )$ around the origin, after computation
in terms of the original mapping $\Phi (y,\lambda ),$ are given below:%
\begin{equation*}
\begin{tabular}{lll}
$g_{x}$ & $=$ & $0,$ \\ 
$g_{xx}$ & $=$ & $\left\langle v_{0}^{\ast },d^{2}\Phi
(v_{0},v_{0})\right\rangle ,$ \\ 
$g_{xxx}$ & $=$ & $\left\langle v_{0}^{\ast },d^{3}\Phi
(v_{0},v_{0},v_{0})-3d^{2}\Phi (v_{0},L^{-1}Ed^{2}\Phi
(v_{0},v_{0}))\right\rangle ,$ \\ 
$g_{\lambda }$ & $=$ & $\left\langle v_{0}^{\ast },\Phi _{\lambda
}\right\rangle ,$ \\ 
$g_{\lambda x}$ & $=$ & $\left\langle v_{0}^{\ast },d\Phi _{\lambda }\cdot
v_{0}-d^{2}\Phi (v_{0},L^{-1}E\Phi _{\lambda }\right\rangle .$%
\end{tabular}%
\end{equation*}%
\medskip

\bigskip

\noindent\textbf{Appendix B. Gravity effects in the model of Holmes}

Solution of equation (\ref{11})%
\begin{equation}
t^{2}w^{\prime \prime }+tw^{\prime }+(t^{2}-(\frac{1}{3})^{2})w=0,  \tag{B.1}
\end{equation}%
may be expressed as%
\begin{equation*}
w(t)=c_{1}\text{J}_{1/3}(t)+c_{2}\text{J}_{-1/3}(t)
\end{equation*}%
so that solution of eq.(11) may be written as%
\begin{eqnarray}
u(t) &=&\frac{\sqrt{\alpha ^{2}+\gamma \xi }}{\gamma ^{1/3}}\left[ c_{1}%
\text{J}_{1/3}\left( \frac{2}{3}\frac{\alpha ^{2}+\gamma \xi )^{3/2}}{\gamma 
}\right) \right.  \notag \\
&&\left. +\text{ }c_{2}\text{J}_{-1/3}\left( \frac{2}{3}\frac{\alpha
^{2}+\gamma \xi )^{3/2}}{\gamma }\right) \right] .  \TCItag{B.2}  \label{b.2}
\end{eqnarray}%
Bessel functions posses the following well known properties, cf. Gradshteyn
and Ryzhik (1994):%
\begin{gather}
\frac{d}{dx}\left( \frac{\text{J}_{\nu }(x)}{x^{\nu }}\right) =-\frac{\text{J%
}_{\nu +1}(x)}{x^{\nu }},  \notag \\
\frac{d}{dx}\left( x^{\nu }\text{J}_{\nu }(x)\right) =x^{\nu }\text{J}_{\nu
-1}(x),  \tag{B.3}  \label{b.3}
\end{gather}%
so that%
\begin{gather}
\int x^{-\nu +1}\text{J}_{\nu }(x)dx=-x^{-\nu +1}\text{J}_{\nu -1}(x), 
\notag \\
\int x^{\text{ }\nu +1}\text{J}_{\nu }(x)dx=\text{ }x^{\text{ }\nu +1}\text{J%
}_{\nu -1}(x).  \tag{B.4}  \label{b.4}
\end{gather}%
Furthermore, boundary conditions of Eq.(11)%
\begin{gather*}
u^{\prime }\mid _{\xi =0}=0 \\
u^{\prime }\mid _{\xi =1}=0,
\end{gather*}%
yield%
\begin{gather}
c_{1}\text{J}_{-2/3}\left( \frac{2}{3}\frac{\alpha ^{3}}{\gamma }\right)
-c_{2}\text{J}_{2/3}\left( \frac{2}{3}\frac{\alpha ^{3}}{\gamma }\right) =0,%
\text{ }  \notag \\
c_{1}\text{J}_{-2/3}\left( \frac{2}{3}\frac{(\alpha ^{2}+\gamma )^{3/2}}{%
\gamma }\right) +c_{2}\text{J}_{2/3}\left( \frac{2}{3}\frac{(\alpha
^{2}+\gamma )^{3/2}}{\gamma }\right) =0.  \tag{B.5}
\end{gather}%
The condition for obtaining nontrivial solutions $c_{1},c_{2}\neq 0$ requires%
\begin{gather}
\text{J}_{2/3}\left( \frac{2}{3}\frac{\alpha ^{3}}{\gamma }\right) \text{J}%
_{-2/3}\left( \frac{2}{3}\frac{(\alpha ^{2}+\gamma )^{3/2}}{\gamma }\right) +
\notag \\
\text{J}_{2/3}\left( \frac{2}{3}\frac{(\alpha ^{2}+\gamma )^{3/2}}{\gamma }%
\right) \text{J}_{-2/3}\left( \frac{2}{3}\frac{\alpha ^{3}}{\gamma }\right)
=0.  \tag{B.6}  \label{b.6}
\end{gather}%
Assuming weak gravitational influence $(\gamma \simeq 0)$, the arguments of
the Bessel functions in the above equation are large so that asymptotic
expressions for Bessel functions may be used%
\begin{gather*}
\text{J}_{\nu }(x)=\sqrt{\frac{2}{\pi x}}\cos (x-\frac{\pi }{2}\nu -\frac{%
\pi }{4})+\vartheta (x^{-3/2}), \\
\text{J}_{-\nu }(x)=\sqrt{\frac{2}{\pi x}}\cos (x+\frac{\pi }{2}\nu -\frac{%
\pi }{4})+\vartheta (x^{-3/2}).
\end{gather*}%
Hence, 
\begin{gather}
\text{J}_{2/3}(x)=\sqrt{\frac{2}{\pi x}}\cos (x-\frac{7\pi }{12}),  \notag \\
\text{J}_{-2/3}(x)=\sqrt{\frac{2}{\pi x}}\cos (x+\frac{7\pi }{12}). 
\tag{B.7}  \label{b.7}
\end{gather}%
In a straightforward manner, condition (\ref{B.6}) yields%
\begin{equation}
\frac{2}{3}\frac{(\alpha ^{2}+\gamma )^{3/2}-\alpha ^{3}}{\gamma }=n\pi . 
\tag{B.8}  \label{B.8}
\end{equation}%
Recalling that $\alpha ^{2}+\gamma =\rho ^{2},$ the above condition is
equivalent to%
\begin{equation*}
\rho ^{3}-(\rho ^{2}-\gamma )^{3/2}=\frac{3}{2}n\pi \gamma ,
\end{equation*}%
which, assuming $\gamma <<\rho $ may be Taylor expanded, and retaining terms
to second order yields%
\begin{equation*}
\rho _{k}=n\pi +\frac{\gamma }{4k\pi }+\vartheta (\gamma ^{2}).
\end{equation*}%
Keeping in mind the substitutions introduced in order to cast the equation
in an analytically solvable form, the solution we seek is%
\begin{equation}
v(\xi )=\int\nolimits_{0}^{\xi }u(\xi )d\xi ,\text{ \ with }u(0)=0. 
\tag{B.9}
\end{equation}%
Inserting eq.(\ref{b.2}) in (\ref{b.3}) the following expression is obtained:%
\begin{equation}
v(x)=\frac{c_{1}}{\gamma ^{1/3}}\left( \frac{3\gamma }{2}\right)
^{2/3}\int\nolimits_{0}^{x}\eta ^{2/3}\text{J}_{1/3}(\eta )d\eta +\frac{c_{2}%
}{\gamma ^{1/3}}\left( \frac{3\gamma }{2}\right)
^{2/3}\int\nolimits_{0}^{x}\eta ^{2/3}\text{J}_{-1/3}(\eta )d\eta , 
\tag{B.10}
\end{equation}%
so that 
\begin{eqnarray}
v(\xi ) &=&\tilde{c}_{1}\left[ (\alpha ^{2}+\gamma \xi )\text{J}%
_{-2/3}\left( \frac{2}{3}\frac{(\alpha ^{2}+\gamma \xi )^{3/2}}{\gamma }%
\right) -\alpha ^{2}\text{J}_{-2/3}\left( \frac{2}{3}\frac{\alpha ^{3}}{%
\gamma }\right) \right]  \notag \\
&&+\tilde{c}_{2}\left[ (\alpha ^{2}+\gamma \xi )\text{J}_{2/3}\left( \frac{2%
}{3}\frac{(\alpha ^{2}+\gamma \xi )^{3/2}}{\gamma }\right) -\alpha ^{2}\text{%
J}_{2/3}\left( \frac{2}{3}\frac{\alpha ^{3}}{\gamma }\right) \right] . 
\TCItag{B.11}
\end{eqnarray}%
Constants in the above two expressions satisfy the following relations:%
\begin{equation}
\frac{c_{2}}{c_{1}}=\frac{\text{J}_{-2/3}\left( \frac{2}{3}\frac{\alpha ^{3}%
}{\gamma }\right) }{\text{J}_{2/3}\left( \frac{2}{3}\frac{\alpha ^{3}}{%
\gamma }\right) }=\frac{\tilde{c}_{2}}{\tilde{c}_{1}}.  \tag{B.12}
\end{equation}%
Finally, the complete solution $v(x)$ may be written as%
\begin{eqnarray}
v(\xi ) &=&\mu (\rho )(\alpha ^{2}+\gamma \xi )\left[ \text{J}_{2/3}\left( 
\frac{2}{3}\frac{\alpha ^{3}}{\gamma }\right) \text{J}_{-2/3}\left( \frac{2}{%
3}\frac{(\alpha ^{2}+\gamma \xi )^{3/2}}{\gamma }\right) \right.  \notag \\
&&\left. -\text{J}_{-2/3}\left( \frac{2}{3}\frac{\alpha ^{3}}{\gamma }%
\right) \text{J}_{2/3}\left( \frac{2}{3}\frac{(\alpha ^{2}+\gamma \xi )^{3/2}%
}{\gamma }\right) +\vartheta (\xi ^{2})\right] ;  \TCItag{B.13}
\end{eqnarray}%
for $\gamma \thickapprox 0,$ this acquires the asymptotic form%
\begin{equation}
u(\xi )\simeq \frac{1}{(\alpha ^{2}+\gamma \xi )^{2}}\sin \left[ \frac{2}{3}%
\frac{(\alpha ^{2}+\gamma \xi )^{3/2}-\alpha ^{3}}{\gamma }\right] , 
\tag{B.14}
\end{equation}%
while 
\begin{equation}
\lim_{\gamma \rightarrow 0}u(\xi )=\sin (n\pi \xi ),  \tag{B.15}
\end{equation}%
as expected.

\noindent\textbf{Appendix C. Normal form of the bifurcation equation for the
complete nonlinear model}

The normal form of the pitchfork bifurcation modulo higher order terms
(h.o.t), reads 
\begin{equation*}
g(\mu ,\rho )=g_{\mu \mu \mu }\mu ^{3}+g_{\mu \rho }\mu \rho +\text{h.o.t.}
\end{equation*}%
As shown in the main part of the paper, explicit determination of the term $%
g_{\mu \rho }$ is not necessary since only its sign is relevant, and it was
demonstrated that it is negative. The remaining term, $g_{\mu \mu \mu }$,
requires explicit determination in order not only to evaluate its sign which
enables classification of bifurcation into supercritical or subcritical
type, but also to extract conditions, in the form of inequalities, that need
to be fulfilled in order for each bifurcation type to arise. In Appendix A,
it was shown that this term requires evaluation of the inner product which
may be written as%
\begin{equation}
g_{xxx}=\left\langle v_{0}^{\ast }|d^{3}\Phi
(v_{0},v_{0},v_{0})\right\rangle -\left\langle v_{0}^{\ast }|3d^{2}\Phi
(v_{0},L^{-1}Ed^{2}\Phi (v_{0},v_{0})\right\rangle .  \tag{C.1}  \label{C.1}
\end{equation}%
Noting that the adjoint operator $L^{\ast }$ of 
\begin{equation*}
L=\left( 
\begin{array}{c}
w_{0}^{\prime \prime \prime \prime }+[(\rho ^{2}-(\Gamma -\Pi ))-\gamma
(1-\xi )]w_{0}^{\prime \prime }+\gamma w_{0}^{\prime } \\ 
u_{0}^{\prime \prime }%
\end{array}%
\right) ,
\end{equation*}%
is equal to $L$, the following expression is obtained for the first inner
product in (\ref{C.1}):%
\begin{gather*}
\left\langle v_{0}^{\ast }|d^{3}\Phi (v_{0},v_{0},v_{0})\right\rangle
=9\alpha _{0}\dint\nolimits_{0}^{1}v_{0}^{\prime 2}v_{0}^{\prime \prime
}v_{0}d\xi +9\gamma \dint\nolimits_{0}^{1}(1-\xi )v_{0}^{\prime
2}v_{0}^{\prime \prime }v_{0}d\xi - \\
\left[ 3\gamma \dint\nolimits_{0}^{1}v_{0}^{\prime 3}v_{0}d\xi
+12\dint\nolimits_{0}^{1}v_{0}^{\prime \prime 3}v_{0}d\xi
+12\dint\nolimits_{0}^{1}v_{0}^{\prime 2}v_{0}^{\prime \prime \prime \prime
}v_{0}d\xi +48\dint\nolimits_{0}^{1}v_{0}v_{0}^{\prime }v_{0}^{\prime \prime
}v_{0}^{\prime \prime \prime }d\xi \right] ,
\end{gather*}%
where%
\begin{equation*}
\alpha _{0}=\Gamma -\mathcal{A}-\Pi ,
\end{equation*}%
and where $v_{0}$ represents the solution of $L=0$ evaluated using procedure
presented in Appendix B. With $\zeta $ given by expression (\ref{48}) $v_{0}$
has the following form%
\begin{align}
v_{0}& =\left( 
\begin{array}{c}
w_{0}(x) \\ 
u_{0}(x)%
\end{array}%
\right) =  \notag \\
& \left( 
\begin{array}{c}
1 \\ 
0%
\end{array}%
\right) (\zeta ^{2}+\gamma \xi )\left[ \text{J}_{2/3}\left( \frac{2}{3}\frac{%
\zeta ^{3}}{\gamma }\right) \text{J}_{-2/3}\left( \frac{2}{3}\frac{(\zeta
^{2}+\gamma \xi )^{3/2}}{\gamma }\right) \right.  \notag \\
& \left. -\text{J}_{-2/3}\left( \frac{2}{3}\frac{\zeta ^{3}}{\gamma }\right) 
\text{J}_{2/3}\left( \frac{2}{3}\frac{(\zeta ^{2}+\gamma \xi )^{3/2}}{\gamma 
}\right) \right] .  \tag{C.2}
\end{align}%
Once the inverse operator $\ L^{-1}Ed^{2}\Phi (v_{0},v_{0})$ is evaluated
following a lengthy procedure, the second inner product requires evaluation
of the following integrals:%
\begin{gather}
\left\langle v_{0}^{\ast }|3d^{2}\Phi (v_{0},L^{-1}Ed^{2}\Phi
(v_{0},v_{0})\right\rangle =  \notag \\
(\alpha _{0}+\gamma )\dint\nolimits_{0}^{1}(\eta ^{\prime }+\eta ^{\prime
\prime })v_{0}v_{0}^{\prime \prime }d\xi -\dint\nolimits_{0}^{1}\xi \eta
^{\prime }v_{0}v_{0}^{\prime \prime }d\xi -  \notag \\
\dint\nolimits_{0}^{1}\xi \eta ^{\prime \prime }v_{0}v_{0}^{\prime }d\xi
-3\dint\nolimits_{0}^{1}\eta ^{\prime \prime \prime }v_{0}v_{0}^{\prime
\prime }-4\dint\nolimits_{0}^{1}\eta ^{\prime \prime }v_{0}v_{0}^{\prime
\prime \prime }d\xi d\xi  \notag \\
-2\dint\nolimits_{0}^{1}\eta ^{\prime }v_{0}v_{0}^{\prime \prime \prime
\prime }d\xi -\dint\nolimits_{0}^{1}\eta ^{\prime \prime \prime \prime
}v_{0}^{\prime }d\xi -\gamma \dint\nolimits_{0}^{1}\eta ^{\prime
}v_{0}v_{0}^{\prime }d\xi ,  \tag{C.3}
\end{gather}%
where%
\begin{gather}
\eta ^{\prime }=\frac{2(\alpha _{0}+\gamma )}{\rho ^{2}-\mathcal{A}}\dint
v_{0}^{\prime }v_{0}^{\prime \prime }d\xi -\frac{2\gamma }{\rho ^{2}-%
\mathcal{A}}\dint \xi v_{0}^{\prime }v_{0}^{\prime \prime }d\xi  \notag \\
-\frac{2}{\rho ^{2}-\mathcal{A}}\dint \left( v_{0}^{\prime }v_{0}^{\prime
\prime \prime \prime }+v_{0}^{\prime \prime }v_{0}^{\prime \prime \prime
}\right) d\xi -\frac{\gamma }{\rho ^{2}-\mathcal{A}}\dint v_{0}^{\prime
2}d\xi .  \tag{C.4}
\end{gather}

Evaluation of integrals was performed using approximate expressions (\ref%
{B.7}). Results were checked using lower and upper bounds for the Bessel
functions, given by the following inequality (Neuman, 2004):%
\begin{gather}
\frac{1}{\Gamma (\alpha +1)}\left( \frac{x}{2}\right) ^{\alpha }\cos \left( 
\frac{x}{\sqrt{2(\alpha +1}}\right) \leq J_{\alpha }  \notag \\
\leq \frac{1}{\Gamma (\alpha +1)}\left( \frac{x}{2}\right) ^{\alpha }\frac{1%
}{3(\alpha +1)}\left[ 2\alpha +1+\left( \alpha +2\right) \cos \left( \sqrt{%
\frac{3}{2\alpha +2}}x\right) \right] .  \tag{C.5}
\end{gather}%
\qquad

After lengthy calculations the following expression is obtained%
\begin{equation}
g_{\mu \mu \mu }=-\left[ \Lambda (\zeta )\frac{(\Gamma -\mathcal{A}-\Pi
+\gamma )}{\rho _{n}^{2}-\mathcal{A}}\left( \frac{2}{3}\frac{\zeta ^{3}}{%
\gamma }\right) +\Delta (\zeta )\frac{1}{\rho _{n}^{2}-\mathcal{A}}\left( 
\frac{2}{3\gamma }\right) ^{1/3}\zeta +\Omega (\zeta )\right] ,  \tag{C.6}
\end{equation}%
where%
\begin{gather}
\Lambda (\zeta )=\left[ a\left( \frac{2}{3}\gamma \right) ^{2}\left( \text{J}%
_{-2/3}\left( \frac{2}{3}\frac{\zeta ^{3}}{\gamma }\right) \right) ^{3}\text{%
J}_{2/3}\left( \frac{2}{3}\frac{\zeta ^{3}}{\gamma }\right) \right.  \notag
\\
\left. +b\left( \frac{2}{3}\gamma \right) ^{8/3}\left( \text{J}_{-2/3}\left( 
\frac{2}{3}\frac{\zeta ^{3}}{\gamma }\right) \right) ^{4}\right] ;  \tag{C.7}
\end{gather}%
$a$ and $b$ are positive constants

\begin{gather*}
a=\frac{1}{\left( \Gamma \left( \frac{2}{3}\right) \right) ^{2}}\left( \frac{%
1}{\Gamma \left( \frac{1}{3}\right) }+\frac{1}{\Gamma \left( \frac{2}{3}%
\right) }\right) ^{2} \\
b=a\left( \frac{1}{\Gamma \left( \frac{1}{3}\right) }+\frac{1}{\Gamma \left( 
\frac{2}{3}\right) }\right) ^{-1}.
\end{gather*}%
Furthermore, term $\Delta (\zeta )$ is 
\begin{eqnarray}
\Delta (\zeta ) &=&d_{1}\left( \frac{2}{3}\right) ^{1/3}\gamma ^{5/3}\left( 
\text{J}_{-2/3}\left( \frac{2}{3}\frac{\zeta ^{3}}{\gamma }\right) \right)
^{3}\text{J}_{2/3}\left( \frac{2}{3}\frac{\zeta ^{3}}{\gamma }\right) + 
\notag \\
&&\left( \frac{2}{3}\gamma \right) ^{10/3}\left( \text{J}_{-2/3}\left( \frac{%
2}{3}\frac{\zeta ^{3}}{\gamma }\right) \right) ^{4}\frac{2^{1/3}}{\left(
\Gamma \left( \frac{2}{3}\right) \right) ^{2}}\left( \frac{4}{39}\right) , 
\TCItag{C.8}
\end{eqnarray}%
where%
\begin{equation*}
d_{1}=\frac{1}{\Gamma \left( \frac{2}{3}\right) }\left( \frac{1}{\Gamma
\left( \frac{1}{3}\right) }+\frac{1}{\Gamma \left( \frac{2}{3}\right) }%
\right) \left( \frac{3}{\Gamma \left( \frac{1}{3}\right) }+\frac{4}{\Gamma
\left( \frac{2}{3}\right) }\right) .
\end{equation*}%
Finally, term $\Omega (\zeta )$ is equal to 
\begin{equation*}
-(\varphi +(\Gamma -\mathcal{A}-\Pi +\gamma )\psi ),
\end{equation*}%
where%
\begin{align*}
\varphi & =-c_{1}\left( \frac{2}{3}\gamma \right) ^{10/3}\left( \text{J}%
_{-2/3}\left( \frac{2}{3}\frac{\zeta ^{3}}{\gamma }\right) \right)
^{2}\left( \text{J}\left( \frac{2}{3}\frac{\zeta ^{3}}{\gamma }\right)
\right) ^{2} \\
& -c_{2}\left( \frac{2}{3}\gamma \right) ^{8/3}\left( \text{J}_{-2/3}\left( 
\frac{2}{3}\frac{\zeta ^{3}}{\gamma }\right) \right) ^{4} \\
& -c_{3}\left( \frac{2}{3}\gamma \right) ^{7/3}\left( \text{J}_{-2/3}\left( 
\frac{2}{3}\frac{\zeta ^{3}}{\gamma }\right) \right) ^{3}\text{J}%
_{2/3}\left( \frac{2}{3}\frac{\zeta ^{3}}{\gamma }\right) , \\
\psi & =\left( \text{J}_{-2/3}\left( \frac{2}{3}\frac{\zeta ^{3}}{\gamma }%
\right) \right) ^{4}\left[ \left( \frac{2}{3}\gamma \right) ^{7/3}+\left( 
\frac{2}{3}\gamma \right) ^{2}\frac{2^{1/3}}{\left( \Gamma \left( \frac{2}{3}%
\right) \right) ^{2}}\right] \left( \frac{1}{\Gamma \left( \frac{1}{3}%
\right) }+\frac{1}{\Gamma \left( \frac{2}{3}\right) }\right) ;
\end{align*}%
$c_{1}$, $c_{2}$ and $c_{3}$ are the following constants:%
\begin{gather*}
c_{1}=12\frac{2^{1/3}}{\Gamma \left( \frac{4}{3}\right) \Gamma \left( \frac{1%
}{3}\right) }\left( \frac{1}{\Gamma \left( \frac{1}{3}\right) }+\frac{1}{%
\Gamma \left( \frac{2}{3}\right) }\right) , \\
c_{2}=\frac{1}{\Gamma \left( \frac{2}{3}\right) }\left( \frac{1}{\Gamma
\left( \frac{1}{3}\right) }-\frac{1}{\Gamma \left( \frac{2}{3}\right) }%
\right) ^{2}, \\
c_{3}=\frac{2^{1/3}}{\Gamma \left( \frac{2}{3}\right) }\left( \frac{1}{%
\Gamma \left( \frac{4}{3}\right) }-\frac{1}{\Gamma \left( \frac{1}{3}\right) 
}\right) .
\end{gather*}%
\newpage

\noindent\textbf{References}

\noindent Argentina, M., Coullet, P., 1998. A generic mechanism for
spatiotemporal intermittency. Physica A 257, 45-60.

\noindent Ch'ng, E., 1977. A theoretical analysis of nonlinear effects on
the flutter and divergence of a tube conveying fluid. Dept. of Mechanical
and Aerospace Engineering, Princeton Univ., AMS report No. 1343 (revised).

\noindent Ch'ng, E., Dowell, E. H., 1979. A theoretical analysis of
nonlinear effects on the flutter and divergence of a tube conveying fluid.
In: Flow-induced Vibrations, Eds. Chen, S. S., Bernstein, M. D.,ASME, New
York, pp. 65-81.

\noindent Gradshteyn, I. S., Ryzhik, I. M., 1994. Table of Integrals, Series
and Products, 5th edition. Academic Press, New York.

\noindent Holmes, P. J., 1977. Bifurcations to divergence and flutter in
flow-induced oscillations: a finite dimensional analysis. Journal of Sound
and Vibration 53, 471 -- 503.

\noindent Holmes, P. J., 1978. Pipes supported at both ends cannot flutter.
Journal of Applied Mechanics 45, 619-622.

\noindent Neuman, E., 2004. Inequalities involving Bessel functions of the
first kind. Journal.of Inequalities in Pure and Applied. Mathematics 5(4),
art. 94.

\noindent Golubitsky, M., Stewart, I., Schaeffer, D. G., 1985. Singularities
and Groups in Bifurcation Theory, Volume I, Springer-Verlag, New York.

\noindent Golubitsky, M., Schaeffer, 1988. D. G., Singularities and Groups
in Bifurcation Theory, Volume II, Springer-Verlag, New York.

\noindent Lunn, T. S., 1982. Flow induced instabilities of fluid-conveying
pipes, Ph.D. Thesis, University College London, London, U.K.

\noindent Paidoussis, M. P., 1998. Fluid-Structure Interactions: Slender
Structures and Axial Flow, Volume 1, Academic Press, London.

\noindent Paidoussis, M. P., 2003. Fluid-Structure Interactions: Slender
Structures and Axial Flow, Volume 2, Elsevier Academic Press, London.

\noindent Rajkovi\'{c}, M., Nikoli\'{c}, M., 2005. Bifurcations in nonlinear
models of fluid-conveying pipes on elastic foundation (in preparation).

\noindent Semler, C., Li, G. X., Paidoussis, M. P., 1994. The nonlinear
equations of motion of pipes conveying fluid. Journal of Fluids and
Structures 10, 787-825.

\noindent Thurman, A. L., Mote, C. D. Jr., 1969. Non-linear oscillations of
a cylinder containing flowing fluid. ASME Journal of Engineering for
Industry 91, 1147-1155.

\bigskip \newpage

\textbf{Figure captions}

Fig. 1. A pipe with supported ends conveying fluid. The diagram also shows
coordinates used in the text; $\rho $ represents the fluid velocity.

\bigskip

Fig. 2. Supercritical pitchfork bifurcation for the model of Holmes with and
without gravity and tensioning effects. Fluid flow is in the same direction
as gravity.

\bigskip

Fig. 3. \ Supercritical pitchfork bifurcation for the model of Holmes. Fluid
flow is in the direction opposite to the direction of gravity.

\bigskip

Fig. 4. \ Subcritical pitchfork bifurcation for the model of Thurman and
Mote.

\bigskip

Fig. 5. Supercritical bifurcation for the fluid conveying pipe with and
without elastic support. Dashed lines correspond to the case without elastic
support. The case with elastic support assumes $K\gg 0,$ and $(1/2)\Gamma
/n\pi \ll n\pi .$ The first three modes are shown.

\bigskip

Fig. A1. A geometrical interpretation of the Lyapunov-Schmidt reduction.

\newpage

\bigskip

\textbf{Tables}

Table 1.

\begin{equation*}
\begin{tabular}{|c|c|c|}
\hline
\textsc{Model} & \multicolumn{2}{|c|}{\textsc{Critical velocity }$\rho _{n}$}
\\ \hline
& ${\small \gamma \geq 0}$ & ${\small \gamma <0}$ \\ \hline
{\small Holmes} & ${\small n\pi +}\frac{1}{4}\frac{\gamma }{n\pi }{\small +}%
\frac{1}{2}\frac{\Gamma }{n\pi }$ & ${\small n\pi -}\frac{1}{4}\frac{\gamma 
}{n\pi }{\small +}\frac{1}{2}\frac{\Gamma }{n\pi }$ \\ \hline
{\small TM and TM-}$\kappa ^{2}${\small \ } \thinspace $(\gamma =0)$ & $%
{\small n\pi +}\frac{1}{2}\frac{\Gamma }{n\pi }$ &  \\ \hline
{\small TM-K} $(\gamma =0)$ & ${\small n\pi +}\frac{1}{2}\frac{\Gamma }{%
(n\pi )}{\small +}\frac{1}{2}\frac{K}{(n\pi )^{3}}$ &  \\ \hline
{\small Complete nonlinear model} & ${\small n\pi +}\frac{1}{4}\frac{\gamma 
}{n\pi }{\small +}\frac{1}{2}\frac{\Gamma }{n\pi }{\small -}\frac{1}{2}\frac{%
\Pi }{n\pi }$ & ${\small n\pi -}\frac{1}{4}\frac{\gamma }{n\pi }{\small +}%
\frac{1}{2}\frac{\Gamma }{n\pi }{\small -}\frac{1}{2}\frac{\Pi }{n\pi }$ \\ 
\hline
\end{tabular}%
\end{equation*}

{\small Table 1. Critical velocity values for models of Holmes, Thurman and
Mote (TM), Thurman and Mote with curvature (TM-}$\kappa ^{2}${\small ) and
the complete nonlinear model of Paidoussis.}

\bigskip

\newpage

Table 2.

\bigskip

\begin{equation*}
\begin{tabular}{|c|c|c|}
\hline
& \multicolumn{2}{|c|}{\textsc{High velocity limit}} \\ \hline
\textsc{Model} & {\small Supercritical condition} & {\small Subcritical
condition.} \\ \hline
{\small TM} & ${\small \Gamma <}\mathcal{A}$ & ${\small \Gamma >}\mathcal{A}$
\\ \hline
{\small TM-}$\kappa ^{2}${\small \ } & ${\small \Gamma <}\mathcal{A}{\small +%
}\frac{10}{3}{\small (n\pi )}^{2}$ & ${\small \Gamma >}\mathcal{A}{\small +}%
\frac{10}{3}{\small (n\pi )}^{2}$ \\ \hline
{\small Complete nonlinear model} & ${\small \Gamma +\gamma <}\mathcal{A}%
{\small +\Pi }$ & ${\small \Gamma +\gamma >}\mathcal{A}{\small +\Pi }$ \\ 
\hline
\end{tabular}%
\end{equation*}

{\small Table 2. Conditions for development of supercritical an subcritical
bifurcations in the high velocity limit}

\newpage

Table 3.\bigskip

\begin{equation*}
\begin{tabular}{|c|c|c|}
\hline
& \multicolumn{2}{|c|}{\textsc{Low velocity limit}} \\ \hline
\textsc{Model} & {\small Supercritical condition} & {\small Subcritical
condition.} \\ \hline
{\small TM} & ${\small \Gamma <}\mathcal{A}{\small -}\frac{1}{2}{\small %
(n\pi )}^{2}$ & ${\small \Gamma >}\mathcal{A}{\small -}\frac{1}{2}{\small %
(n\pi )}^{2}$ \\ \hline
{\small TM-}$\kappa ^{2}${\small \ } & ${\small \Gamma <}\mathcal{A}{\small +%
}\frac{7}{4}{\small (n\pi )}^{2}$ & ${\small \Gamma >}\mathcal{A}{\small +}%
\frac{7}{4}{\small (n\pi )}^{2}$ \\ \hline
{\small Complete nonlinear model} & ${\small \Gamma +\gamma >}\mathcal{A}%
{\small +\Pi +}\frac{\left\vert \varphi \right\vert }{\psi }$ & ${\small %
\Gamma +\gamma <}\mathcal{A}{\small +\Pi +}\frac{\left\vert \varphi
\right\vert }{\psi }{\small .}$ \\ \hline
\end{tabular}%
\end{equation*}

{\small Table 3. Conditions for development of supercritical and subcritical
bifurcations in different models of fluid conveying pipes. Term }$\left\vert
\varphi \right\vert {\small /\psi ,}${\small proportional to }$\gamma ,$ 
{\small is defined in eq.(\ref{66}).}

\end{document}